\documentclass[fleqn,usenatbib]{mnras}

\usepackage{newtxtext,newtxmath}
\usepackage[T1]{fontenc}
\usepackage{booktabs}

\usepackage{amssymb}
\usepackage{tabularx, siunitx}
\usepackage[flushleft]{threeparttable}
\usepackage[normalem]{ulem}
\usepackage{enumitem}

\DeclareRobustCommand{\VAN}[3]{#2}
\let\VANthebibliography\thebibliography
\def\thebibliography{\DeclareRobustCommand{\VAN}[3]{##3}\VANthebibliography}

\usepackage{graphicx}	
\usepackage{amsmath}	
\usepackage{multirow}   
\usepackage{newtxmath}
\usepackage{rotating}
\usepackage{pdflscape}
\usepackage{bm}

\title[The Ionospheric Cut-off as a Potential Technosignature]
{The Ionospheric Cut-off as a Potential Radio Technosignature}

\author[M.A. Garrett]{Michael A. Garrett$^{1,2,3}$
\\
$^{1}$Jodrell Bank Centre for Astrophysics, Department of Physics and Astronomy, Alan Turing Building, University of Manchester, Oxford Road, M13 9PL, UK\\
$^{2}$Leiden Observatory, Leiden University, P.O. Box 9513, NL-2300 RA Leiden, The Netherlands\\
$^{3}$University of Malta, Institute of Space Sciences and Astronomy, Msida, MSD2080, Malta\\
}

\pubyear{2026}

\begin{document}

\label{firstpage}
\pagerange{\pageref{firstpage}--\pageref{lastpage}}
\maketitle


\begin{abstract}

Radio emission generated on a planetary surface must cross the planet's ionosphere before escaping into space: below the local plasma frequency it is reflected back towards the surface. We investigate this process for broadband radio technosignatures (BRaTs) using a Chapman-layer description of photoionised atmospheres. The plasma frequency depends only weakly on stellar XUV irradiation, scaling as $Q_{\infty}^{1/4}$, implying turnover frequencies of 5--40~MHz for most terrestrial habitable-zone planets, potentially rising towards 100~MHz in extreme environments. The basic signature is simple: the continuum of a surface-origin BRaT should be strongly suppressed below a characteristic frequency that can be estimated from the XUV output of the host star, at every orbital phase and for any viewing geometry. Oblique propagation, magnetic fields, and collisional absorption broaden the cut-off into a smooth escape function, $f_{\rm esc}(\nu,\phi)$, where $\phi$ is the orbital phase. Because the day--night ionisation pattern is anchored to the substellar point rather than to the rotating surface, disc integration adds a second, more discriminating signature: the half-escape frequency $\nu_{1/2}$ varies predictably (by $\approx 40$--$50$ per cent over the orbit, depending on the angular emission pattern of the leakage) and remains locked to the orbital ephemeris. Finite atmospheric response times introduce modest phase lags but preserve the modulation. A low-frequency turnover alone is not unique to technology, but one that varies periodically with the orbital period would be difficult to reproduce through known astrophysical processes. Although observationally challenging, the ionospheric cut-off is a potential technosignature diagnostic and a novel probe of exoplanet ionospheres.

\end{abstract}



\section{Introduction}
\label{sec:intro}

For over sixty years, radio SETI has prioritised narrowband, largely monochromatic emissions, a strategy driven by their energetic efficiency and the absence of natural confounding sources. While this approach remains highly relevant \citep{CocconiMorrison1959,Tarter2001,enriquez_breakthrough_2017,price_breakthrough_2020}, the parameter space for potential technosignatures is significantly broader. Recent work has begun to explore the complementary regime of \emph{broadband} radio technosignatures (BRaTs), in which the integrated, unintentional electromagnetic leakage of a planetary-scale technological society produces a measurable enhancement in the radio continuum of an unresolved exoplanet \citep{2026MNRAS.549ag991G}. BRaTs offer a new and largely unexplored search domain. 

A central question for any BRaTs search programme is the choice of observing frequency. On Earth, the ground-accessible radio window is bounded at low frequencies by the terrestrial ionosphere, near $\sim 10$~MHz, and at high frequencies by increasing atmospheric and molecular absorption beyond 50~GHz \citep{Wilson2013}. The same physics, with very different parameters, governs the escape of artificial radio emission from the surface of an exoplanet into space.  Below the local plasma frequency, the electromagnetic wave is reflected by the ionosphere and ultimately dissipated within the atmosphere. Well above the local plasma frequency, the wave escapes into space with little impedance. As emphasised by \citet{2026MNRAS.549ag991G}, the implication for SETI can be highly significant: low-frequency leakage from worlds whose ionospheres are appreciably denser than Earth's may be entirely invisible to a distant observer.

This paper expands on the possible impact of the ionospheric cut-off on BRaTs as noted by \cite{2026MNRAS.549ag991G}. It develops the ionospheric cut-off as both a constraint and a possible signature for BRaTs. Our motivation is threefold. First, the physical conditions governing the ionospheric cut-off on exoplanets remain poorly explored, especially for terrestrial worlds in the habitable zones of low-mass, active stars. Secondly, the cut-off is in itself a simple and testable prediction: radio emission that originates at a planetary surface cannot be seen below the plasma frequency of the overlying ionosphere, so the continuum spectrum of a BRaT should turn over at a frequency that can be estimated in advance from the XUV output of the host star. This holds at every orbital phase, and testing it requires no knowledge of the orbit. Finally, we argue that because the cut-off frequency itself varies in a deterministic, geometry-locked manner, it offers a diagnostic that known natural sources are unlikely to mimic in full. Natural processes can produce low-frequency spectral turnovers, and some, notably auroral cyclotron-maser emission and star–planet interactions, can produce periodic variability (Section~\ref{conf}). However, no known natural process presents the specific combination that the ionospheric cut-off predicts: a smooth, weakly polarised turnover whose characteristic frequency varies with planetary phase. The detection of a broadband planetary radio continuum exhibiting a low-frequency turnover, whose frequency oscillates in phase with the planet's orbital geometry, would therefore provide strong supporting evidence for a planetary-surface broadband radio technosignature. 

The paper is organised as follows. Section~\ref{sec:physics} summarises the plasma physics of the ionospheric cut-off, including the effects of the planetary magnetic field and oblique propagation. Section~\ref{sec:ss} reviews ionospheric properties across the Solar
System as empirical benchmarks, and Section~\ref{sec:exo} builds on this with models of the expected cut-off frequencies of terrestrial exoplanet ionospheres as a function of their stellar host parameters. Section~\ref{sec:diurnal} develops
the central result: the orbital-phase modulation of the apparent ionospheric frequency cut-off, which an unresolved external observer would measure, and the natural processes that could mimic it. Section~\ref{sec:obs} outlines an observational strategy and the criteria a candidate detection should satisfy. Section~\ref{sec:discussion} discusses the broader value of the cut-off as a planetary atmospheric diagnostic, and we draw our conclusions in
Section~\ref{sec:conc}.

\section{Ionospheric plasma physics}
\label{sec:physics}

\subsection{The plasma frequency}

Consider first an idealised ionosphere: a cold, unmagnetised plasma with a smooth electron-density profile $n_e$ that varies only with altitude, free of horizontal gradients or irregularities (collisions and magnetic fields are introduced below). The response of the free electrons to an externally applied electromagnetic wave gives rise to the plasma frequency \citep{RybickiLightman1986,Kelley2009}
\begin{equation}
\nu_p = \frac{1}{2\pi} \sqrt{\frac{n_e e^2}{m_e \epsilon_0}} \approx 8.98 \sqrt{n_e/\mathrm{m^{-3}}} \;\;\mathrm{Hz},
\label{eq:nu_p}
\end{equation}
where $e$ is the elementary charge, $m_e$ the electron mass, and $\epsilon_0$ the vacuum permittivity. Electromagnetic waves with angular frequency $\omega < \omega_p \equiv 2\pi\nu_p$ have an imaginary wavenumber and are reflected. Waves with $\omega > \omega_p$ propagate with a refractive index
\begin{equation}
n(\omega) = \sqrt{1 - (\omega_p/\omega)^2},
\label{eq:n_index}
\end{equation}
approaching unity at high frequencies.

For an inhomogeneous ionosphere stratified primarily in altitude, radio-wave propagation is determined by the electron-density profile $n_e(h)$ rather than by altitude itself. Altitude matters only because the electron density, and hence the local plasma frequency, vary with height. A wave of frequency $\nu$ can propagate only through regions where $\nu > \nu_p(h)$. Reflection occurs at the turning point where the local plasma frequency first becomes equal to the wave frequency, $\nu_p(h)=\nu$. The maximum electron density encountered along the propagation path therefore determines whether a wave can escape into space. We refer to the corresponding frequency as the \emph{critical frequency} $\nu_c$, which for vertical incidence is equal to the plasma frequency at the density peak of the ionospheric layer.

\subsection{Oblique propagation}

A surface-based transmitter emits radiation into the upward hemisphere, spanning a range of propagation directions. While vertically propagating waves encounter the minimum ionospheric column density, obliquely propagating rays traverse a longer path through the stratified plasma.

For vertical propagation, electromagnetic waves escape provided their frequency exceeds the local ionospheric cut-off, $\nu > \nu_{c}$.
For a ray propagating at an angle $\theta$ to the local vertical, the tangential component of the wave vector is conserved in a horizontally stratified medium; it is the normal component that determines whether the ray continues upward or reaches a turning point. In the plane-parallel approximation, the component of the incident wave vector normal to the ionospheric layers is reduced by a factor $\cos\theta$, yielding the escape criterion,
\begin{equation}
\nu\cos\theta > \nu_c,
\qquad\text{or equivalently}\qquad
\nu_{\rm esc}(\theta)=\frac{\nu_c}{\cos\theta}.
\label{eq:sec}
\end{equation}


A consequence of this secant-law dependence is that emission at frequencies modestly above $\nu_c$ can still be trapped if launched at sufficiently low elevation. For a planetary-scale leakage source, the limb-grazing components are systematically suppressed by the obliquity factor, broadening the spectral turnover. The broadening is of order unity in $\nu / \nu_c$ for a broad, quasi-isotropic distribution of launch angles, although the precise factor depends on the transmitters' angular radiation patterns, on ionospheric curvature and refraction, and on the quantity being observed. The ionospheric cut-off is therefore smoother than an abrupt break at a single, well-defined frequency.

Consequently, the minimum frequency required for escape increases with the propagation angle. Rays directed close to the horizon $ (\theta\rightarrow90^\circ)$  require progressively higher frequencies to penetrate the ionosphere, while exactly horizontal propagation does not lead to escape within the plane-parallel approximation. This formal divergence is an artefact of the plane-parallel approximation: for a real spherical planet the ionosphere is curved, so a ray that is initially horizontal with respect to the surface does not remain parallel to every overlying layer, and its actual trajectory is determined by planetary curvature, the finite thickness of the ionosphere, and refraction. A simple regularisation, which we use in Section~\ref{sec:discavg}, treats the ionosphere as a thin spherical shell at height $h$ above a planet of radius $R$. A ray leaving the surface at zenith angle $\theta$ then crosses the shell at the smaller angle $\theta'$, with $\sin\theta' = [R/(R+h)]\sin\theta$, so that the obliquity factor becomes
\[
S(\mu) \;=\; \Bigl[\,1 - \Bigl(\tfrac{R}{R+h}\Bigr)^{2}(1-\mu^{2})\Bigr]^{-1/2},
\qquad \mu \equiv \cos\theta ,
\]
which reproduces the secant law $1/\mu$ for small $\theta$ but remains finite at the limb, $S_{\max} = [1-(R/(R+h))^{2}]^{-1/2} \approx 3.4$ for $h/R \approx 0.047$ (Earth's F2 peak). Refraction and the finite thickness of real layers modify this further, but the thin-shell form captures the essential regularisation.

\subsection{Magnetised plasma effects}

So far, we have treated the ionosphere as an unmagnetised plasma. This is a useful first approximation, but real planets often have magnetic fields. In a magnetic field, free electrons do not simply oscillate back and forth in response to a radio wave. They also spiral around the magnetic field lines. This changes the way radio waves propagate through the plasma \citep{Budden1985}.

The characteristic frequency of this spiral motion is called the electron gyrofrequency:
\begin{equation}
\nu_B = \frac{eB}{2\pi m_e}
       \approx 28\,(B/\mu\mathrm{T})\;\mathrm{kHz}.
\label{eq:nu_B}
\end{equation}
Here \(B\) is the magnetic-field strength, \(e\) is the electron charge, and \(m_e\) is the electron mass. For Earth, \(B \approx 50\,\mu\mathrm{T}\), so
\[
\nu_B \approx 1.4~\mathrm{MHz}.
\]
This frequency is significantly lower than the daytime ionospheric plasma cut-off, which can approach approximately \(10\)~MHz (the peak plasma frequency varies considerably with location, local time, season, and solar activity), but it is not entirely negligible.

The magnetic field splits the radio-wave propagation into two natural modes, usually called the ordinary mode (O-mode) and the extraordinary mode (X-mode). These modes behave slightly differently because the electrons are forced to move in circular paths around the magnetic field. For waves travelling perpendicular to the magnetic field, the upper X-mode cut-off frequency is
\begin{equation}
\nu_X =
\frac{1}{2}
\left[
\nu_B +
\left(\nu_B^2 + 4\nu_p^2\right)^{1/2}
\right].
\label{eq:nu_X}
\end{equation}
If the plasma frequency is much larger than the gyrofrequency, \(\nu_p \gg \nu_B\), this becomes approximately
\[
\nu_X \approx \nu_p + \frac{\nu_B}{2}.
\]
This is usually a good approximation for Earth-like ionospheres. For propagation parallel to the field the natural modes are instead the right- and left-hand circularly polarised (R and L) modes; equation~(\ref{eq:nu_X}) then gives the R-mode cut-off, while the L-mode cut-off, $\nu_L = \tfrac{1}{2}[-\nu_B + (\nu_B^2 + 4\nu_p^2)^{1/2}]$, lies below $\nu_p$.

The main point is that a magnetic field splits the cut-off rather than simply shifting it: the X-mode cut-off lies above $\nu_p$, the O-mode cut-off remains near $\nu_p$, and other mode boundaries can lie below it. Instead of a single cutoff frequency, the ionosphere has different cut-offs for different wave modes and directions of travel. The exact answer depends on the angle between the wave direction and the magnetic field, as well as the local plasma density and magnetic field strength. 
Again, the overall effect is to smear out the spectral turnover rather than to produce a perfectly clean edge.

Electron–neutral collisions also provide an additional source of attenuation near the plasma cut-off \citep{Kelley2009}. Collisions do not alter the formal plasma cut-off at $\nu_p$; rather, they introduce absorption that broadens the transition, so that appreciable power may escape only at frequencies somewhat above the nominal cut-off. This effect is strongest in dense, weakly ionised ionospheres or lower atmospheric plasma layers, where the neutral density, and hence the electron--neutral collision frequency, is high. A large electron density alone does not produce strong collisional absorption.

\subsection{The Chapman layer and time variability}

The ionosphere is produced when ultraviolet and X-ray radiation from a star ionises the upper atmosphere. The electron density is therefore not uniform with altitude. At very high altitudes, there are relatively few neutral atoms available to ionise, while at lower altitudes, much of the ionising radiation has already been absorbed. As a result, the electron density reaches a maximum at an intermediate height.

A simple and widely used description of this structure is the Chapman layer \citep{Chapman1931,Kelley2009}. In this model, the equilibrium electron density at altitude \(h\) is
\begin{equation}
n_e(h) = n_e^{\max}
\exp\!\left[
\frac{1}{2}
\left(
1-z-\sec\chi\,e^{-z}
\right)
\right],
\label{eq:chapman}
\end{equation}
where \(z=(h-h_{\max})/H\) is the altitude in units of the atmospheric scale height \(H\), and \(h_{\max}\) and \(n_e^{\max}\) in this expression are the height and density of the peak for an overhead star (\(\chi=0\)). At zenith angle \(\chi\) the peak rises by \(H\ln\sec\chi\) and its density falls as \((\cos\chi)^{1/2}\), consistent with the peak scaling derived below. The angle \(\chi\) is the stellar zenith angle, i.e. the angle between the local vertical and the direction to the host star. At local noon \(\chi=0\), while near sunrise and sunset \(\chi\) approaches \(90^\circ\).

Although equation~(\ref{eq:chapman}) describes the detailed vertical structure of the ionosphere, the quantity most relevant for radio-wave escape is the peak electron density, \(n_e^{\max}\), since the plasma cut-off is determined by the densest part of the layer.

The scaling of \(n_e^{\max}\) can be understood from a simple ionisation--recombination balance \citep[the \(\alpha\)-Chapman equilibrium; e.g.][]{Chapman1931,Schunk2009}. In equilibrium, the rate at which ionising photons create free electrons must be balanced by the rate at which electrons and ions recombine. If the ionisation rate is proportional to the incident photon flux, \(Q_\infty \cos\chi\), and the recombination rate is proportional to \(\alpha n_e^2\), then
\begin{equation}
\label{eq:balance}
Q_\infty \cos\chi \sim \alpha n_e^2.
\end{equation}
Here \(Q_\infty\) is the ionising photon flux at the top of the atmosphere and \(\alpha\) is the recombination coefficient of the dominant ion species. Solving for the equilibrium electron density gives
\begin{equation}
n_e^{\max}
\propto
\sqrt{\frac{Q_\infty \cos\chi}{\alpha}}.
\end{equation}
Since the plasma frequency scales as \(\nu_p \propto \sqrt{n_e}\), the maximum plasma frequency in the ionosphere follows
\begin{equation}
\nu_p^{\max}
\propto
Q_\infty^{1/4}
(\cos\chi)^{1/4}
\alpha^{-1/4}.
\label{eq:scaling}
\end{equation}

Equations~(\ref{eq:balance})--(\ref{eq:scaling}) are proportionality relations: dimensionless atmospheric and geometric factors -- the neutral scale height $H$, the ionisation efficiency, and the Chapman-function geometry -- have been absorbed into the constants of proportionality, or equivalently into an effective $Q_\infty$. They therefore describe scalings among atmospheres of broadly similar structure; strong differences in scale height or composition enter through these omitted factors, a point we return to in Section~\ref{sec:exo}.

This quarter-power dependence is an important scaling factor that plays a central role in the discussion of exoplanet ionospheres (Section~\ref{sec:exo}) and the phase modulation of the cut-off frequency (Section~\ref{sec:diurnal}). Because the plasma frequency depends only on the fourth root of the ionising flux, even large changes in stellar irradiation produce relatively modest changes in the cut-off frequency. For example, doubling the cut-off frequency requires a sixteen-fold increase in the ionising flux. Similarly, the variation in \(\cos\chi\) across the dayside causes the ionospheric electron density, and hence the plasma cut-off frequency, to vary predictably between local noon and the day-night terminator. However, changes in the cut-off frequency are only a factor of a few between local noon and the day–night terminator, rather than orders-of-magnitude variations.

\section{Ionospheres of the Solar System}
\label{sec:ss}

The Solar System provides empirical anchors that span more than three orders of magnitude in peak electron density and provide empirical anchors for much of the parameter space relevant to BRaTs (Table~\ref{tab:ss}). Earth's mid-latitude F2 layer reaches $n_e^{\max} \sim 10^{12}~\mathrm{m^{-3}}$ at solar maximum and during the day, dropping by an order of magnitude at night and during solar minimum \citep{Kelley2009}. Venus, despite receiving roughly twice the solar extreme-ultraviolet (EUV) flux of Earth (inverse-square scaling at 0.72~au), has a comparable cut-off because its dense CO$_2$ atmosphere recombines more efficiently - it supports a dayside ionosphere with $n_e^{\max} \sim 6\times10^{11}~\mathrm{m^{-3}}$ near 140~km altitude \citep{Brace1979,Bauer2004}. Mars exhibits a tenuous ionosphere with $n_e^{\max} \sim 1.7\times10^{11}~\mathrm{m^{-3}}$ at 130~km \citep{Withers2009}.

\begin{table}
\centering
\caption{Representative ionospheric peak electron densities and corresponding vertical plasma cut-off frequencies in the Solar System.}
\label{tab:ss}
\begin{tabular}{lccc}
\hline
Body & $n_e^{\max}$ (m$^{-3}$) & $\nu_p$ (MHz) & Ref. \\
\hline
Mercury & $\sim 10^{9}$ & $\sim 0.3$ & a \\
Venus (day) & $6\times10^{11}$ & 7 & b \\
Earth (day, max) & $1$--$2\times10^{12}$ & 9--13 & c \\
Earth (night) & $1$--$3\times10^{11}$ & 3--5 & c \\
Mars (day) & $1.7\times10^{11}$ & 4 & d \\
Jupiter & $\sim 2\times10^{11}$ & 4 & e \\
Saturn & $\sim 10^{10}$--$10^{11}$ & 1--3 & e \\
Titan & $\sim 3\times10^{9}$ & 0.5 & f \\
Uranus & $\sim 10^{10}$ & 1 & e \\
Neptune & $\sim 10^{10}$ & 1 & e \\
\hline
\end{tabular}
\\[2pt]
{\footnotesize \emph{References:} (a) \citet{Russell2007Mercury}; (b) \citet{Brace1979}; (c) \citet{Kelley2009}; (d) \citet{Withers2009}; (e) \citet{Yelle2004}; (f) \citet{Wahlund2005}.}
\end{table}

Several features of Table~\ref{tab:ss} are worth noting. First, the variation among bodies, despite differences of orders of magnitude in stellar irradiation, atmospheric composition, and magnetic environment, spans a factor of $\sim 2000$ in $n_e^{\max}$, from Mercury to Earth's daytime maximum -- equivalent to a factor of $\sim 45$ in $\nu_p$. This range, modest compared with the $\sim 10^{4}$ spread in solar irradiation alone across the same bodies, partly reflects the quarter-power scaling of equation~(\ref{eq:scaling}) and partly the fact that, in steady state, recombination and ionisation balance limit excursions in $n_e$. Secondly, the night-side of a predominantly photoionised ionosphere typically exhibits a significantly lower cut-off than the day side, by a factor of typically a few, providing the empirical basis for the day--night modulation argument developed later. This contrast is a common tendency rather than a universal rule: it depends on recombination, dayside-to-nightside transport, particle precipitation, auroral processes, and magnetospheric plasma sources, any of which can sustain a substantial nightside density.
Thirdly, despite their large sizes, the giant planets have only weakly ionised upper atmospheres: relatively few free electrons are produced by solar EUV radiation at these distances, and those that are produced are rapidly removed by dissociative recombination with molecular ions such as H$_3^+$, a process that neutralises the ion and yields neutral fragments.

\section{Exoplanet ionospheric cut-offs}
\label{sec:exo}

The discussion of Section~\ref{sec:physics} establishes that the radio cut-off of a planetary
ionosphere is set by the peak electron density at the layer maximum, which, in turn, is set by the
balance between stellar photoionisation and ionic recombination (equation~\ref{eq:scaling}). To
predict the cut-off frequency for a given exoplanet, we need to know how much ionising
radiation the planet receives at the top of its atmosphere and how efficiently its dominant ions
recombine.

\subsection{Stellar EUV and X-ray driving}

The photons responsible for ionising planetary atmospheres lie almost entirely in the extreme-ultraviolet and X-ray bands (XUV). We adopt the conventional integration boundary $\lambda \lesssim 912$~\AA{} ($E \gtrsim 13.6$~eV), which corresponds to the ionisation threshold of atomic hydrogen; it is a bookkeeping convention rather than a universal photoionisation threshold, since several molecular constituents ionise at lower energies (e.g.\ $\approx 12.1$~eV for O$_2$ and $\approx 12.6$~eV for H$_2$O), so that the relevant band is composition-dependent. At these energies, the photons can liberate electrons from the dominant neutral constituents (N$_2$, O$_2$, CO$_2$, H$_2$O), and the strength of the XUV flux at the top of the atmosphere directly sets the primary ionisation rate and hence the ionospheric electron density \citep{France2013,Peacock2019}.

Two features of low-mass stars matter here. First, even in their quiescent state, M-dwarfs emit a larger \emph{fraction} of their bolometric output in the XUV than Sun-like stars do, often by several orders of magnitude in the most active cases. Second, the habitable zones of M-dwarfs lie much closer to the host star than the Sun's habitable zone, mainly because M-dwarfs have low bolometric luminosities. Both effects compound, so a habitable-zone planet orbiting an M-dwarf receives XUV flux many times that of Earth's. 

A convenient parameterisation is the photon flux at the top of the planetary atmosphere,
\begin{equation}
Q_\infty \;=\; \frac{L_{\rm XUV}}{4\pi a^{2}\,\langle E_\gamma\rangle},
\label{eq:Qinf}
\end{equation}
where $L_{\rm XUV}$ is the stellar XUV luminosity, $a$ is the orbital separation, and
$\langle E_\gamma\rangle$ is the mean photon energy in the relevant band. For Earth,
$Q_\oplus \approx 3\times10^{14}$~photons~m$^{-2}$~s$^{-1}$ \citep[e.g.][]{Schunk2009}. Proxima Centauri~b \citep{2016Natur.536..437A}, in the habitable
zone of a moderately active M5.5V star, receives an XUV irradiance $\approx 60$ times that of Earth \citep{Ribas2017}; assuming a mean XUV photon energy comparable to the solar value, this corresponds to $Q_\infty/Q_\oplus \approx 60$. The early M-dwarf AD~Leo (M3V) places its habitable zone in a still harsher XUV
environment, with $Q_\infty/Q_\oplus \approx 200$ (estimated from the quiescent XUV spectrum adopted by \citealt{2010AsBio..10..751S}). The seven known terrestrial planets of
TRAPPIST-1 \citep{2017Natur.542..456G} span a more modest $Q_\infty/Q_\oplus \approx 1$--$15$ in quiescence, although the
system is also known for strong, frequent flares \citep{Wheatley2017,Vida2017}.

\subsection{Scaling the cut-off frequency to other planets}

Combining the Chapman scaling of equation~(\ref{eq:scaling}) with the plasma-frequency relation
of equation~(\ref{eq:nu_p}), and normalising to Earth's daytime peak ($\nu_c \approx 10$~MHz at
$Q_\oplus$), we obtain
\begin{equation}
\nu_c \;\approx\; 10~{\rm MHz}\,
\left(\frac{Q_\infty}{Q_\oplus}\right)^{1/4}\,
\left(\frac{\alpha}{\alpha_\oplus}\right)^{-1/4}.
\label{eq:nu_c_scaling}
\end{equation}
Equation~(\ref{eq:nu_c_scaling}) implicitly assumes that the atmospheric scale height, ionisation efficiency, illumination geometry, and XUV spectral shape are Earth-like (cf.\ Section~\ref{sec:physics}); it should be read as a scaling among broadly terrestrial atmospheres rather than as a general formula.
Here $\alpha_\oplus \approx 3\times10^{-13}$~m$^{3}$~s$^{-1}$ is a representative laboratory dissociative-recombination coefficient for the molecular ions NO$^{+}$ and O$_2^{+}$ at $T_e \sim 1000$~K \citep{Schunk2009}, adopted as the loss coefficient in the $\alpha n_e^{2}$ closure of equation~(\ref{eq:balance}). We emphasise what this coefficient does and does not represent. The terrestrial F2-region peak is dominated by atomic O$^{+}$, which is removed through ion--neutral reactions that produce molecular ions (NO$^{+}$ and O$_2^{+}$); these molecular ions then undergo rapid dissociative recombination (molecular ions can replace O$^{+}$ as the dominant F-region ion under particular disturbed or high-latitude conditions; e.g.\ \citealt{banks1974no}). The first, ion--neutral step is generally rate-limiting and is linear rather than quadratic in $n_e$, so a phenomenological coefficient that reproduced the combined sequence would depend on neutral density, composition, altitude, temperature, and $n_e$ itself, and would not equal the molecular-ion dissociative-recombination coefficient. We therefore keep the two quantities distinct: the ratio $\alpha/\alpha_\oplus$ in equation~(\ref{eq:nu_c_scaling}) compares laboratory dissociative-recombination coefficients of the dominant molecular ions across atmospheric chemistries -- a defined, if simplified, comparison -- whereas the empirical atmospheric relaxation time invoked in Sections~\ref{sec:flares} and \ref{sec:ode} is a separate parameter, introduced and calibrated there. A CO$_2$-dominated atmosphere recombines through O$_2^{+}$
with $\alpha \sim 2\times10^{-13}$~m$^{3}$~s$^{-1}$; an H$_2$-dominated atmosphere (e.g.\ a
retained primordial envelope on a sub-Neptune) recombines mainly through H$_3^{+}$ with
$\alpha \sim 10^{-13}\text{--}10^{-12}$~m$^{3}$~s$^{-1}$, depending on electron temperature and the degree of internal excitation. Because the dependence is fourth-root, the cut-off is
only weakly sensitive to uncertainty in a fixed effective recombination coefficient, although substantially different atmospheric chemistries may also alter the scale height and the dominant loss processes.

\subsection{Representative cases}

Table~\ref{tab:exo} applies equation~(\ref{eq:nu_c_scaling}) to a small set of representative
habitable-zone targets. For an Earth-analogue at 1~AU around a Sun-like star (G2V), we recover
$\nu_c \approx 10$~MHz by construction. The quiescent late-G dwarf $\tau$~Ceti (G8V) hosts the candidate planet
$\tau$~Ceti~e in a slightly weaker XUV environment than Earth, with $\nu_c \approx 9$~MHz. A hypothetical terrestrial planet in the habitable zone of the young, active K-dwarf $\epsilon$~Eridani would receive $Q_\infty/Q_\oplus \approx 15$, giving
$\nu_c \approx 20$~MHz (the known planet $\epsilon$~Eri~b is a $\approx 1\,M_{\rm Jup}$ giant on a near-circular orbit at $\approx 3.5$~au \citep{Thompson2025} and is not itself a BRaTs target). For terrestrial planets around active M-dwarfs, the cut-off climbs into the
tens of MHz: Proxima~b sits at $\nu_c \approx 28$~MHz, and a planet in AD~Leo's habitable zone at
$\nu_c \approx 38$~MHz. 

\begin{table}
\centering
\caption{Estimated quiescent and flare-state plasma cut-off frequencies for representative
habitable-zone targets, computed from equation~(\ref{eq:nu_c_scaling}) with Earth-like atmospheric
composition. The flare column applies equation~(\ref{eq:nu_c_scaling}) with the target-dependent sustained XUV enhancement factor listed in the fourth column: a modest $3\times$ for the Sun-like and late-G hosts, $10\times$ for the active K dwarf, and $100\times$ -- a fiducial strong-flare scenario -- for the M dwarfs. Because $\nu_c \propto Q_\infty^{1/4}$, even a $10^{3}\times$ enhancement raises $\nu_c$
by only a factor of $\sim 5.6$. The flare values are equilibrium responses, attained only if the enhancement is sustained for longer than the ionospheric response time (see Section~\ref{sec:flares}).}
\label{tab:exo}
\begin{tabular}{lcccc}
\hline
Target & $Q_\infty/Q_\oplus$ & $\nu_c$ quiet & XUV enh. & $\nu_c$ flare \\
       &                     & (MHz)         & (factor) & (MHz) \\
\hline
Sun--Earth                   & 1     & 10  & 3   & 13  \\
$\tau$~Ceti e (G8V)          & 0.6   & 9   & 3   & 12  \\
$\epsilon$~Eri HZ (K2V)$^{a}$  & 15    & 20  & 10  & 36  \\
TRAPPIST-1e (M8V)            & 2     & 12  & 100 & 38  \\
Proxima~b (M5.5V)            & 60    & 28  & 100 & 89  \\
AD~Leo HZ (M3V)              & 200   & 38  & 100 & 120 \\
\hline
\end{tabular}
\\[2pt]
{\footnotesize
$^a$A hypothetical terrestrial planet in the habitable zone of $\epsilon$~Eridani; the known planet $\epsilon$~Eri~b is a $\sim1\,M_{\rm Jup}$ giant at $\approx3.5$~au \citep{Thompson2025}.}
\end{table}

A few general remarks follow from Table~\ref{tab:exo}. For terrestrial planets in the habitable
zones of Sun-like and K-dwarf hosts, the cut-off remains close to Earth's value, in the
5--30~MHz range; low-frequency BRaTs leakage from such systems is in principle accessible to a
sufficiently sensitive metre-wave observatory. For active M-dwarfs, the cut-off climbs into the
tens of MHz, potentially rising towards $\sim100$~MHz in the most extreme flare-driven or strongly irradiated environments, which significantly restricts the accessible window below 1~GHz but
does not close it. In all cases, observing at $\gtrsim 1$~GHz remains a safe choice for
detection-driven BRaTs searches that are willing to forego the diagnostic value of the cut-off
itself (Section~\ref{sec:diurnal}).

\subsection{Flares and the recombination floor}
\label{sec:flares}

The numbers in the third column of Table~\ref{tab:exo} refer to the quiescent state of each star.
M-dwarfs, however, are well known to produce frequent, energetic flares with optical energies typically
$10^{32}$--$10^{34}$~erg \citep{Howard2019}. The conversion from optical flare energy to instantaneous XUV enhancement is uncertain and flare-dependent, but enhancements of factors $\sim 10$--$10^{3}$ above the quiescent XUV level have been inferred for strong events, on timescales of minutes to hours; we therefore treat the enhancement factor as a free parameter in what follows.
Through equation~(\ref{eq:nu_c_scaling}) a $100\times$ XUV enhancement raises $\nu_c$ by a factor
of $\sim 3.2$; a $1000\times$ enhancement by $\sim 5.6$. The flare column of
Table~\ref{tab:exo} applies target-dependent sustained enhancements -- $3\times$ for the Sun-like and late-G hosts, $10\times$ for the active K dwarf, and the fiducial strong-flare scenario of $100\times$ for the M dwarfs -- giving transient cut-offs of $\approx 89$~MHz for Proxima~b and $\approx 120$~MHz for a planet in AD~Leo's habitable zone. Reaching the gigahertz regime through flaring is effectively impossible: because $\nu_c \propto Q_\infty^{1/4}$, raising the cut-off from 28~MHz (Proxima~b) or 38~MHz (AD~Leo) to 1~GHz would require XUV enhancements of $(1000/28)^{4} \approx 1.6\times10^{6}$ and $(1000/38)^{4} \approx 5\times10^{5}$ relative to quiescence, orders of magnitude beyond any observed flare.

These flare-state values are equilibrium responses: they assume the ionosphere has time to come into a new Chapman equilibrium with the elevated XUV flux. In practice, the time required is set by the recombination timescale. For a simple ionosphere dominated by one positive ion species, electron loss occurs through recombination,
\begin{equation*}
e^- + X^+ \rightarrow X ,
\end{equation*}
so the loss rate is proportional to the product of the electron and ion number densities,
\begin{equation*}
\left(\frac{dn_e}{dt}\right)_{\rm loss} = -\alpha n_e n_i ,
\end{equation*}
where $\alpha$ is the effective recombination coefficient. Since the plasma is quasi-neutral on
the relevant scales, $n_i \simeq n_e$, and hence
\begin{equation*}
\dot{n}_e \equiv \frac{dn_e}{dt} \simeq -\alpha n_e^{2},
\end{equation*}
which is the loss term in equation~(\ref{eq:balance}). A characteristic recombination time is then
obtained by dividing the electron density by its loss rate
\citep[e.g.][]{Bauer2004,Schunk2009},
\begin{equation}
\tau_{\rm rec} \;\sim\; \frac{n_e}{|\dot{n}_e|}
\;\sim\; \frac{1}{\alpha\,n_e^{\max}},
\label{eq:tau_rec}
\end{equation}
evaluated at the peak of the ionospheric layer. For the daytime Earth-calibrated values adopted here ($\nu_c \approx 10$~MHz, i.e.\ $n_e^{\max} \approx 1.2\times10^{12}$~m$^{-3}$, with $\alpha_\oplus = 3\times10^{-13}$~m$^{3}$~s$^{-1}$), equation~(\ref{eq:tau_rec}) evaluates to $\tau_{\rm rec} \approx 3$~s, rising to tens of seconds at typical night-side densities; it should be regarded as a lower bound on the true response time. In Earth's F region the effective relaxation time is far longer -- tens of minutes to hours -- because the peak is dominated by atomic O$^{+}$, whose removal is limited by the slower ion--neutral chemistry described in Section~\ref{sec:exo}, and because transport contributes to the balance; in that regime the simple $\alpha n_e^{2}$ model is no longer sufficient. Whether a flare drives the cut-off to its equilibrium value therefore depends on which response time applies: bursts shorter than the relevant response time produce only partial enhancements. The flare entries in Table~\ref{tab:exo} should accordingly be read as values attained only in sustained events: a brief flare burst leaves the ionosphere largely undisturbed, while an elevated XUV state lasting longer than the atmospheric response time can drive the cut-off close to the tabulated value.

\subsection{The cut-off as a first-order signature}
\label{sec:static}

\begin{figure}
\centering
\includegraphics[width=\columnwidth]{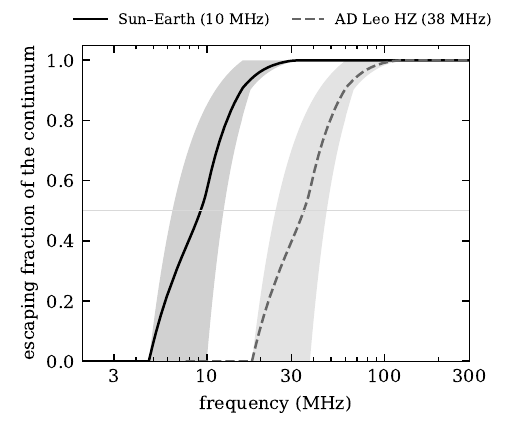}
\caption{The phase-independent turnover. The fraction of surface-origin continuum emission that escapes to a distant observer is shown as a function of frequency for two targets of Table~\ref{tab:exo}: the Sun--Earth case ($\nu_c^{\rm noon} = 10$~MHz) and a planet in the habitable zone of AD~Leo ($\nu_c^{\rm noon} = 38$~MHz). The lines show the orbit-averaged escape fraction of the fiducial disc-integrated model developed in Section~\ref{sec:discavg} (equation~\ref{eq:fesc}); the shaded bands show the range covered over the orbit, from the transit (nightside) spectrum on the left to the eclipse (dayside) spectrum on the right. The turnover moves within the band but is present at every phase. The change of slope near the half-escape level is a consequence of the sharp per-element step adopted in the model, and would be smoothed in a real ionosphere (see Fig.~\ref{fig:modulation}).}
\label{fig:spectrum}
\end{figure}

Before considering any time dependence, we note the simplest observational consequence of the preceding sections. Radio emission generated at the surface of a planet cannot escape below the local plasma frequency: it is reflected back towards the surface. Earth is a familiar example - short-wave broadcasting relies on exactly this reflection. The spectrum of any surface-origin/surface-dominated BRaT must therefore turn over at low frequencies, at a characteristic frequency that equation~(\ref{eq:nu_c_scaling}) allows us to estimate in advance from the stellar XUV flux, for a broadly terrestrial atmosphere: roughly $10$--$40$~MHz for the quiescent targets of Table~\ref{tab:exo}. Seen from a distance, the continuum is progressively suppressed over a band comparable in width to the cut-off frequency itself; for an Earth-like ionosphere, the disc-integrated model of Section~\ref{sec:discavg} places the decline between roughly 16 and 5~MHz (Fig.~\ref{fig:spectrum}).

This turnover is present at every orbital phase and for every viewing geometry, including the face-on orbits for which the phase modulation of Section~\ref{sec:diurnal} vanishes. Its detection requires only a time-averaged, multi-band spectrum of a known BRaT, with no knowledge of the orbital ephemeris. A low-frequency turnover is not unique to this mechanism (Section~\ref{conf}), so it is a necessary rather than a sufficient condition for a surface origin. It is nevertheless an interesting test: continuum emission that persists well below the predicted plasma frequency would be very difficult to reconcile with a surface origin, and would point instead to a source above the ionosphere. The phase dependence developed in Section~\ref{sec:diurnal} refines this basic signature into a more discriminating one.

\section{Orbital-phase modulation as a diagnostic}
\label{sec:diurnal}

\begin{figure}
\centering
\includegraphics[width=\columnwidth]{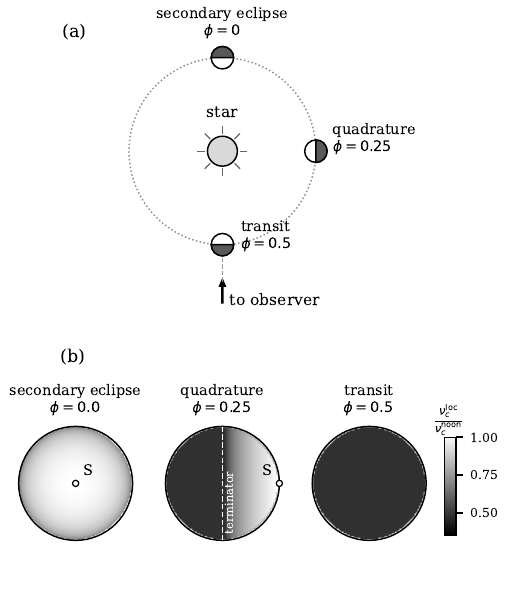}
\caption{Viewing geometry of the phase-locked cut-off.
\textit{Panel (a):} the orbit viewed from above the orbital plane, with the
observer towards the bottom. At secondary eclipse ($\phi=0$) the dayside
hemisphere faces the observer; at transit ($\phi=0.5$) the nightside faces
the observer; quadrature ($\phi=0.25$) presents half of each.
\textit{Panel (b):} the visible disc as seen by the observer at the same
three phases, shaded by the local cut-off
$\nu_c^{\rm loc}/\nu_c^{\rm noon}=[\max(\cos\chi,\xi)]^{1/4}$
(equation~\ref{eq:nu_c_t}).  The greyscale represents the local ionospheric cut-off, not reflected-light brightness. The substellar point (S) sits at the disc centre at
eclipse but is projected onto the limb at quadrature, where the terminator
(dashed) crosses the disc centre; at transit, the disc sits uniformly at the
nightside floor $\xi^{1/4}\approx0.47$ (the uniform shade reflects the
single-valued floor $\xi$, which subsumes dusk-to-dawn structure into one
number; Section~\ref{sec:ode} relaxes this assumption). The visible disc
thus spans the full range of local cut-offs at quadrature, but a single value
at transit -- which is why the disc-integrated turnover is broadest at
quadrature and narrowest at transit (though not sharp: once the dependence of the escape threshold on the emission angle towards the observer is included, Section~\ref{sec:discavg}, elements near the limb escape only at higher frequencies at every phase).}
\label{fig:geometry}
\end{figure}

Section~\ref{sec:exo} treated the ionospheric cut-off as a single number for each planet, set by the quiescent XUV flux of the host star and modified by flaring activity. In reality, the cut-off is not static in space or time. The dayside of the planet is illuminated by its host star; the nightside is not, and the local cut-off at any given location varies systematically with local time. Crucially, however, the resulting day--night pattern is anchored to the substellar point, not to the rotating surface: the surface rotates \emph{through} a pattern that stays fixed relative to the star (on a synchronously rotating planet, by contrast, the pattern is fixed to the surface itself, and the surface does not rotate through it; see Section~\ref{sec:localcut}). What a distant observer can detect is therefore not the rotation of the planet but the changing viewing geometry of this star-anchored pattern, which repeats on the orbital period. Throughout this section we assume leakage that is approximately steady in time and broadly distributed in longitude, and an ionosphere controlled primarily by stellar illumination. Departures from these assumptions -- localised or beamed transmitters, surface geography, magnetically organised ionospheric structure, and atmospheric response lags -- can introduce additional rotation-period signals, and are discussed in Sections~\ref{sec:ode} and \ref{sec:caveats}. Figure~\ref{fig:geometry} presents the geometry of the star-planet-observer system. This section develops that observation into a candidate diagnostic for distinguishing artificial radio emission from natural sources.

\subsection{The local cut-off varies through the day}
\label{sec:localcut}

At any given longitude on a rotating planet, the angle between the local vertical and the direction to the host star --- the stellar zenith angle $\chi$ --- varies sinusoidally with rotation,
\begin{equation}
\cos\chi(t)=\cos\varphi\,\cos\delta\,\cos[\Omega(t-t_0)]+\sin\varphi\,\sin\delta
\label{eq:cos_chi}
\end{equation}
where $\varphi$ is the latitude of the observed longitude, $\delta$ is the stellar declination set by the planet's axial tilt, $\Omega$ is the apparent angular rate of the star in the planet's sky (the solar-day frequency: for a prograde circular orbit, $\Omega = \Omega_{\rm rot} - 2\pi/P_{\rm orb}$, where $\Omega_{\rm rot}$ is the inertial rotation rate); the corresponding solar-day period, $P_{\rm sol} = 2\pi/|\Omega|$, coincides with the rotation period only when $P_{\rm rot} \ll P_{\rm orb}$, and lengthens without bound as synchronous rotation is approached, and $t_0$ is the time of local noon. At $\chi=0$ the star is directly overhead; at $\chi=90^\circ$ the star is on the horizon; on the nightside $\cos\chi$ formally goes negative.

Through the Chapman scaling of equation~(\ref{eq:scaling}), the local plasma cut-off would naively follow $\nu_c^\mathrm{loc} \propto (\cos\chi)^{1/4}$. The Chapman expression assumes photochemical equilibrium between ionisation by the incident stellar XUV and recombination of the dominant ions, and it breaks down on the nightside, where the source term has switched off. There the electron density persists because recombination is not instantaneous, supplemented by plasma transport from the dayside and residual ionisation by cosmic rays and energetic particles \citep{Kelley2009,Schunk2009}. None of these processes are captured by the Chapman expression. We follow a standard phenomenological prescription and introduce a small floor $\xi$ on $\cos\chi$ to prevent the cut-off falling to zero on the nightside,
\begin{equation}
\nu_c^\mathrm{loc}(t)=\nu_c^\mathrm{noon}\,\bigl[\max(\cos\chi(t),\,\xi)\bigr]^{1/4}.
\label{eq:nu_c_t}
\end{equation}
The value of $\xi$ is set empirically. For Earth's F2 layer, the observed daytime peak plasma frequency is $\nu_c^\mathrm{day}\approx 9$--$13$~MHz, falling to $\nu_c^\mathrm{night}\approx 3$--$5$~MHz on the nightside \citep[Table~\ref{tab:ss};][Ch.~5]{Kelley2009}, giving a measured night-to-day cut-off ratio of $\nu_c^\mathrm{night}/\nu_c^\mathrm{noon}\approx 0.3$--$0.5$. Matching the upper end of this range, which corresponds to a well-developed solar-maximum nighttime ionosphere, fixes $\xi\approx 0.05$ (so that $\xi^{1/4}\approx 0.47$). We adopt this value as our Earth-calibrated baseline throughout; in practice, the choice of $\xi$ rolls together all of the non-photoionisation processes that maintain the nightside ionosphere into a single number, and a quantitatively accurate model would replace it with an explicit recombination-and-transport calculation (see also Section~\ref{sec:ode}).

With this prescription, $\nu_c^\mathrm{loc}$ swings between $\nu_c^\mathrm{noon}$ at local noon and $\xi^{1/4}\nu_c^\mathrm{noon}\approx 0.47\,\nu_c^\mathrm{noon}$ on the nightside --- a peak-to-trough modulation of just over 50 per cent at the level of a single longitude in the Earth-calibrated model used here. The fourth-root smooths the variation considerably: even when $\cos\chi$ swings over a factor of twenty between noon and the terminator, $\nu_c^\mathrm{loc}$ varies only by a factor of two.

An important consequence follows immediately. Because equation~(\ref{eq:nu_c_t}) depends on position only through the stellar zenith angle, the steady-state pattern of the local cut-off is a function of local time alone; it is stationary in the frame that co-rotates with the planet–star line. A longitudinally uniform distribution of leakage sources rotating through this stationary pattern therefore produces \emph{no} modulation on the rotation period - rotating a uniform source distribution beneath a star-anchored filter changes nothing. The observable modulation instead arises as the observer's line of sight scans across the pattern with changing orbital geometry (Section~\ref{sec:discavg}). The period and phase of this idealised ionospheric pattern are set entirely by the orbital ephemeris and the system geometry, independent of the source; this geometric phase-locking is the key point. The amplitude and detectability of the observed modulation, by contrast, do still depend on the source -- on its spectrum, spatial distribution, antenna patterns, and duty cycle (Sections~\ref{sec:obs} and \ref{sec:caveats}). Genuine rotation-period signals require structure anchored to the surface or to the planetary magnetic field: most obviously a geographically concentrated transmitter distribution, as on present-day Earth, but natural, magnetically organised ionospheric or magnetospheric structure can also rotate with the planet. Rotational signals are separable from the orbital signal by timescale, but are not by themselves evidence of technology (Section~\ref{sec:caveats}).

For tidally locked planets in the habitable zones of M-dwarfs, the substellar point is, in addition, fixed in the planet's co-rotating frame, so the pattern is static in both frames: the illumination is time-independent at each surface location, and the equilibrium description of equation~(\ref{eq:nu_c_t}) then applies without the rotational lag of Section~\ref{sec:ode} (transport, magnetically organised structure, and flare-driven variability of course remain). For Proxima~b the relevant period is the 11.2-d orbital period. For transiting systems, the variation is locked to the transit and occultation geometry, which is independently determined and provides a stringent cross-check. For non-transiting planets -- including Proxima~b itself, for which photometric searches have found no transits \citep{Kipping2017} -- the orbital period and phase are known from radial velocities, but the inclination, and hence the expected modulation amplitude, remains uncertain (Section~\ref{sec:discavg}).

\subsection{Disc averaging: what a distant observer measures}
\label{sec:discavg}

An external observer on Earth sees a distant exoplanet as an unresolved point source and does not measure the local cut-off discussed in Section~\ref{sec:localcut}. They measure a single integrated spectrum from the whole visible hemisphere of the planet, in which the dayside (with its high cut-off) and the nightside (with its much lower cut-off) blend together, weighted by the projected area of each surface element. The apparent cut-off they extract is therefore an average over the visible disc, and it varies because the visible hemisphere scans across the star-anchored pattern as the planet moves around its orbit: the observer sees the full dayside at superior conjunction, the full nightside at inferior conjunction, and a mixture of the two in between.

The observable is developed in two steps. We first define a simple geometric proxy -- the projected-area-weighted mean of the vertical-incidence cut-off over the visible disc -- which admits closed-form limits and fixes the phase behaviour. We then construct the actual observable, the disc-integrated escape fraction, which additionally incorporates the dependence of the escape threshold on the emission angle towards the observer (Section~\ref{sec:physics}) and an explicit model for the angular emission pattern of the sources. As we show, the second step changes the level and shape of the turnover substantially, but not its fractional modulation.

A convenient proxy for this apparent cut-off is the
projected-area-weighted average of the local cut-off over the visible disc.
Writing $\hat{\bm{n}}$ for the local surface normal, $\hat{\bm{o}}$ for the
unit vector towards the observer, and $\phi$ for the orbital phase
($\phi=0$ at superior conjunction), we define
\begin{equation}
\langle \nu_c \rangle(\phi) \;=\;
\frac{\displaystyle\int_{\hat{\bm{n}}\cdot\hat{\bm{o}}>0}
      \nu_c^{\rm loc}\!\left[\chi(\hat{\bm{n}};\phi)\right]
      \,(\hat{\bm{n}}\cdot\hat{\bm{o}})\;{\rm d}A}
     {\displaystyle\int_{\hat{\bm{n}}\cdot\hat{\bm{o}}>0}
      (\hat{\bm{n}}\cdot\hat{\bm{o}})\;{\rm d}A},
\label{eq:discavg}
\end{equation}
where $\nu_c^{\rm loc}$ is given by equation~(\ref{eq:nu_c_t}) and the
integrals run over the observer-facing hemisphere. Equation~(\ref{eq:discavg}) should be regarded as a geometric proxy for the observable turnover frequency rather than a full radiative-transfer solution. It assumes that
each surface element contributes in proportion to its projected area, and it assigns each element its \emph{vertical-incidence} cut-off, ignoring the fact that the radiation reaching the observer leaves each element obliquely, at emission angle $\cos\theta_o = \mu \equiv \hat{\bm{n}}\cdot\hat{\bm{o}}$, and must therefore satisfy the higher threshold $\nu_c^{\rm loc}\,S(\mu)$ of Section~\ref{sec:physics}. Both omissions are repaired in the escape-fraction model below; the proxy is retained because its closed-form limits make the phase behaviour transparent.

\begin{figure}
\centering
\includegraphics[width=\columnwidth]{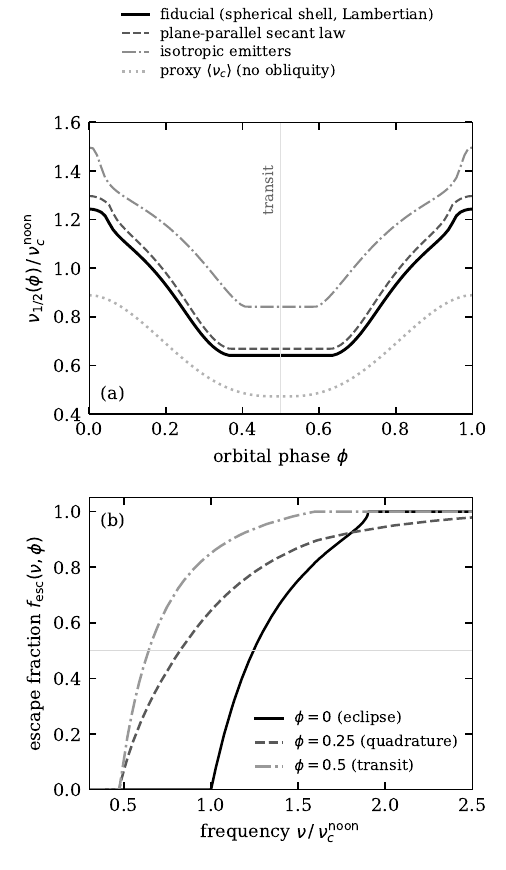}
\caption{Disc-integrated escape model including the observer-directed escape threshold (equation~\ref{eq:fesc}), for the Earth-calibrated nightside floor $\xi = 0.05$, zero obliquity, an edge-on orbit, and uniform intrinsic leakage.
\textit{Panel (a):} the half-escape frequency $\nu_{1/2}(\phi)$ as a function of orbital phase ($\phi = 0$ at superior conjunction). Solid: the fiducial model (thin-shell obliquity factor with $h/R = 0.047$ and Lambertian emission weighting $w = \mu$). Dashed: the plane-parallel secant law $S = 1/\mu$, for which the eclipse and transit values are analytic, $2^{3/8} \approx 1.30$ and $\sqrt{2}\,\xi^{1/4} \approx 0.67$. Dot--dashed: isotropic emitters ($w = 1$). Dotted: the vertical-incidence proxy $\langle\nu_c\rangle$ of equation~(\ref{eq:discavg}), which spans $8/9$ to $\xi^{1/4}$. Including the observer-directed threshold raises the level of $\nu_{1/2}$ by $\approx 40$ per cent and produces a flat plateau at the nightside value for $|\phi - 0.5| \lesssim 0.13$, but leaves the fractional modulation $A = (\nu_{1/2}^{\max} - \nu_{1/2}^{\min})/\nu_{1/2}^{\max}$ essentially unchanged: 48 per cent for the fiducial and plane-parallel models, against 47 per cent for the proxy; isotropic and horizon-peaked emission give 44 and 39 per cent.
\textit{Panel (b):} the escape fraction $f_{\rm esc}(\nu,\phi)$ of the fiducial model at eclipse, quadrature and transit, with a sharp per-element step; the horizontal line marks $f_{\rm esc} = \tfrac{1}{2}$. At eclipse no leakage escapes below $\nu_c^{\rm noon}$, because every element must satisfy $\nu \ge \nu_c^{\rm loc}S(\mu) \ge \nu_c^{\rm noon}$; the extended high-frequency tails arise from elements near the limb. The sharp onset at $\nu_c^{\rm noon}$ and the steep rise near $1.9\,\nu_c^{\rm noon}$ in the eclipse curve arise from the sharp
per-element step and the infinitely thin shell: the observer-directed
threshold peaks at $\approx 1.90\,\nu_c^{\rm noon}$ near $\mu \approx 0.18$
and falls again towards the limb. In a real ionosphere these features would be rounded off by finite layer thickness, refraction, magnetoionic splitting (of order $\nu_B/2$), and collisional absorption (Section~\ref{sec:physics}), with little effect on $\nu_{1/2}$.}
\label{fig:modulation}
\end{figure}

For an edge-on (transiting) geometry, two limits bound the behaviour of equation~(\ref{eq:discavg}), and both can be evaluated in closed form (at lower inclinations the fully illuminated and fully dark hemispheres are never presented to the observer, and these extremes are not reached; see below). When the substellar point faces the observer (the secondary-eclipse geometry of a transiting system) the directions to the star and to the observer coincide, so for every surface element the projection weight equals the cosine of the stellar zenith angle,
$\hat{\bm{n}}\cdot\hat{\bm{o}} = \cos\chi$. Writing $\mu = \cos\chi$ and
noting that the element of surface area on the visible hemisphere is
${\rm d}A \propto {\rm d}\mu$, equation~(\ref{eq:discavg}) with
$\nu_c^{\rm loc} = \nu_c^{\rm noon}\,\mu^{1/4}$ reduces to a one-dimensional
integral,
\begin{equation}
\frac{\langle \nu_c \rangle(0)}{\nu_c^{\rm noon}}
= \frac{\displaystyle\int_0^1 \mu^{1/4}\,\mu\;{\rm d}\mu}
       {\displaystyle\int_0^1 \mu\;{\rm d}\mu}
= \frac{4/9}{1/2}
= \frac{8}{9}
\approx 0.89 ,
\label{eq:eightninths}
\end{equation}
where the factor $\mu^{1/4}$ is the Chapman scaling of the local cut-off and
the additional factor of $\mu$ is the projection weight. 
The apparent cut-off at this phase is thus dragged only slightly below the noon value by the dimmer limb regions seen in morning and evening light. At the opposite extreme, when the antistellar
hemisphere faces the observer (the transit geometry), every element of the
visible disc sits at the nightside floor of equation~(\ref{eq:nu_c_t}), and
the average is exact and trivial:
$\langle \nu_c \rangle(0.5) = \xi^{1/4}\,\nu_c^{\rm noon} \approx
0.47\,\nu_c^{\rm noon}$. The proxy is the dotted curve in
Fig.~\ref{fig:modulation}(a): it glides
smoothly between these two limits as the planet moves around its orbit, with a fractional modulation of $1 - \tfrac{9}{8}\xi^{1/4} \approx 47$ per cent.
The three principal viewing geometries -- secondary
eclipse, quadrature, and transit -- and the corresponding appearance of the
visible disc in terms of the local cut-off $\nu_c^{\rm loc}/\nu_c^{\rm noon}$ are also illustrated in Figure~\ref{fig:geometry}.

The observable itself is the disc-integrated escape fraction. Let $\mu \equiv \hat{\bm{n}}\cdot\hat{\bm{o}}$ be the cosine of the emission angle towards the observer. Radiation from a surface element reaches the observer only above the observer-directed threshold $\nu_c^{\rm loc}\,S(\mu)$, where $S(\mu)$ is the obliquity factor of Section~\ref{sec:physics}: the secant law $S = 1/\mu$ in the plane-parallel limit, or its thin-shell regularisation. The received contribution of each element also depends on the angular emission pattern of the sources, which we describe by a weighting $w(\mu)$ per unit surface area. Then
\begin{equation}
f_{\rm esc}(\nu,\phi)\;=\;
\frac{\displaystyle\int_{\mu>0}
      w(\mu)\;T\!\left[\nu;\; \nu_c^{\rm loc}(\hat{\bm{n}};\phi)\,S(\mu)\right]
      \;{\rm d}A}
     {\displaystyle\int_{\mu>0}
      w(\mu)\;{\rm d}A},
\label{eq:fesc}
\end{equation}
is the fraction of the disc-integrated leakage that escapes to the observer at frequency $\nu$ and phase $\phi$, where $T$ is the per-element transmission profile, a step (or smoothed step) at the observer-directed threshold. Three emission weightings are considered. \emph{Lambertian}, $w = \mu$, is appropriate to a diffuse surface of uniform specific intensity and is adopted as the fiducial case; we stress that this is an assumption, not a generic consequence of uniformly distributed transmitters. \emph{Isotropic}, $w = 1$, describes a uniform population of discrete transmitters each radiating isotropically into the upper hemisphere. \emph{Horizon-peaked}, $w = (1-\mu^{2})^{1/2}$, represents antenna patterns directed towards the horizon, as is common for terrestrial broadcast transmitters. In full generality $T$ depends also on the local magnetic-field geometry and polarisation, $T = T(\nu;\,\nu_c^{\rm loc},\,\mu,\,\bm{B},\,p)$, and a non-uniform intrinsic surface brightness $I_\nu$ would enter the integrand as an additional weight that cancels only in the uniform case (Section~\ref{sec:caveats}). We adopt the scalar, uniform-leakage form of equation~(\ref{eq:fesc}) with a sharp per-element step, so that the analytic limits below can be checked exactly and the effect of obliquity isolated; magnetoionic and collisional broadening would further smooth the resulting curves. We characterise the turnover by the half-escape frequency $\nu_{1/2}(\phi)$, defined by $f_{\rm esc}(\nu_{1/2},\phi) = \tfrac{1}{2}$, and define the fractional modulation as $A \equiv (\nu_{1/2}^{\max} - \nu_{1/2}^{\min})/\nu_{1/2}^{\max}$.

Two limits can again be evaluated in closed form, for the plane-parallel secant law with Lambertian weighting. At superior conjunction $\mu = \cos\chi$, so the threshold is $\nu_c^{\rm noon}\mu^{1/4}/\mu = \nu_c^{\rm noon}\mu^{-3/4}$; with $x \equiv \nu/\nu_c^{\rm noon}$, an element escapes if $\mu \ge x^{-4/3}$, and integrating the projected-area weight $\mu\,{\rm d}\mu$ gives
\begin{align*}
f_{\rm esc}(x,0) &= 1 - x^{-8/3}, \\
\nu_{1/2}(0) &= 2^{3/8}\,\nu_c^{\rm noon} \approx 1.30\,\nu_c^{\rm noon}.
\end{align*}
At transit every element sits at the floor, so the threshold is $\xi^{1/4}\nu_c^{\rm noon}/\mu$ and
\begin{align*}
f_{\rm esc}(x,0.5) &= 1 - \xi^{1/2}\,x^{-2}, \\
\nu_{1/2}(0.5) &= \sqrt{2}\,\xi^{1/4}\,\nu_c^{\rm noon} \approx 0.67\,\nu_c^{\rm noon}.
\end{align*}
Comparison with the proxy limits $8/9$ and $\xi^{1/4}$ shows that the observer-directed threshold raises the level of $\nu_{1/2}$ by $\approx 40$ per cent at both extremes -- a shift of the turnover, not merely a smoothing of its edge -- whereas the fractional modulation, $A = 1 - 2^{1/8}\xi^{1/4} \approx 48$ per cent, is almost identical to the proxy value of 47 per cent, because obliquity multiplies both extremes by similar factors. With the thin-shell regularisation of Section~\ref{sec:physics} ($h/R = 0.047$) the extremes become $1.24$ and $0.64\,\nu_c^{\rm noon}$, again with $A = 48$ per cent (Fig.~\ref{fig:modulation}a). The amplitude does depend on the assumed emission pattern: isotropic emitters give $\nu_{1/2}$ spanning $0.84$--$1.50\,\nu_c^{\rm noon}$ and $A = 44$ per cent, and horizon-peaked emitters $0.97$--$1.59\,\nu_c^{\rm noon}$ and $A = 39$ per cent, because patterns that favour the limb sample more oblique, higher-threshold paths. We therefore quote $A \approx 48$ per cent for the fiducial model and $39$--$48$ per cent across the emission models considered, and regard the emission weighting as an explicit model assumption to be fitted or marginalised over in any real analysis.

Two further features of Fig.~\ref{fig:modulation} follow directly from the geometry. First, $\nu_{1/2}(\phi)$ sits on a flat plateau at its transit value for $|\phi - 0.5| \lesssim 0.13$ (for the fiducial Lambertian weighting; the plateau is narrower for emission patterns that favour the limb): whenever the substellar point lies more than $135^\circ$ from the observer direction, the inner $45^\circ$ of the disc -- half of the projected area -- is entirely on the nightside, so the half-escape frequency is pinned at the nightside value regardless of the dayside sliver at the limb. Secondly, the low-frequency onset of escape is not fixed near the floor at all phases: at superior conjunction no leakage escapes below $\nu_c^{\rm noon}$, since every element must satisfy $\nu \ge \nu_c^{\rm loc}S(\mu) \ge \nu_c^{\rm noon}$, whereas at transit escape begins at $\xi^{1/4}\nu_c^{\rm noon}$ from the disc centre. It is the half-escape frequency, rather than the onset, that we use to characterise the migrating turnover.


The shape of the transition also varies with phase in a predictable
way. Measured between the 10 and 90 per cent escape levels, the fiducial transition spans $0.7$, $1.1$ and $0.6\,\nu_c^{\rm noon}$ at eclipse, quadrature and transit respectively: it is broadest near quadrature, where the visible disc spans the
full range of local cut-offs, and narrowest at the transit geometry,
where the disc sits uniformly at the nightside floor
(Figure~\ref{fig:geometry}) -- though not sharp, since elements near the limb escape only above $\xi^{1/4}S(\mu)\,\nu_c^{\rm noon}$, and every phase carries a high-frequency tail contributed by the limb. This phase-dependent width is itself part
of the template, and provides an additional consistency check beyond
the migration of $\nu_{1/2}$ alone. The family of curves is also
significantly asymmetric: the quadrature spectrum resembles the transit spectrum much more than the eclipse spectrum ($\nu_{1/2} = 0.82$, $0.64$ and $1.24\,\nu_c^{\rm noon}$ respectively). This is because the substellar region is projected onto the limb while the more highly
weighted terminator region occupies the disc centre (Figure~\ref{fig:geometry}, middle panel of b).

The vertical-incidence proxy of equation~(\ref{eq:discavg}) therefore underestimates the observable $\nu_{1/2}$ by $\approx 40$ per cent at all phases and is not itself the template; it is retained only as the geometric skeleton whose closed-form limits explain the phase behaviour. Fig.~\ref{fig:modulation}(a) shows $\nu_{1/2}(\phi)$ itself, and it is this curve -- computed with an explicitly stated obliquity factor and emission weighting -- that constitutes the template light curve used in Section~\ref{sec:obs}.

The fiducial geometry of Figure~\ref{fig:modulation} assumes an edge-on view, in which the substellar and antistellar hemispheres each move fully
into the line of sight. For lower orbital inclinations, the observer never
sees the full day--night contrast: the modulation amplitude is reduced,
vanishing entirely in the face-on limit of this idealised circular, zero-obliquity, axisymmetric model (eccentricity, obliquity and seasonal effects, magnetic asymmetries, or time-variable stellar activity could still induce orbital variability, though not with the specific phase relationship predicted here). This geometric dependence favours an observational programme (see Section~\ref{sec:obs}) that targets close to edge-on transiting systems that present the maximum modulation amplitude and the best-determined ephemerides. For a transiting system the planet is itself occulted at exact superior conjunction, so the full-dayside spectrum of Fig.~\ref{fig:modulation} is sampled immediately before and after secondary eclipse rather than at $\phi = 0$ exactly; the plateau around transit is unaffected, since the planet is then in front of the star.

Finally, we note what survives when the phase information is discarded altogether. Averaged uniformly over the orbit, the fiducial escape fraction has a half-escape frequency of $0.92\,\nu_c^{\rm noon}$ and rises from 10 to 90 per cent between $0.52$ and $1.56\,\nu_c^{\rm noon}$, with no escape below the nightside floor, $\xi^{1/4}\nu_c^{\rm noon} = 0.47\,\nu_c^{\rm noon}$; the orbit-averaged half-escape frequency is $1.14$ and $1.27\,\nu_c^{\rm noon}$ for isotropic and horizon-peaked emitters respectively. For $\nu_c^{\rm noon} = 10$~MHz the fiducial continuum therefore declines between about 16 and 5~MHz. This is the phase-independent turnover of Section~\ref{sec:static}: it is present in any time-averaged spectrum, at any inclination.

\subsection{Departures from equilibrium: rotation and the dusk--dawn asymmetry}
\label{sec:ode}

The equilibrium prescription of equation~(\ref{eq:nu_c_t}) assumes that each
surface element adjusts instantaneously to its local ionising flux. In
practice, the response takes a finite time (Section~\ref{sec:flares}): the electron density of a surface parcel obeys
\begin{equation}
\frac{{\rm d}n_e}{{\rm d}t} \;=\; q(\chi) \;-\; \alpha n_e^{2},
\qquad
q(\chi) \;=\; \alpha \bigl(n_e^{\rm noon}\bigr)^{2}
\max(\cos\chi,\,\xi),
\label{eq:ode}
\end{equation}
where the loss term is that of equation~(\ref{eq:balance}) and $\xi$ now
enters as a \emph{production} floor representing plasma transport and residual night-time ionisation: a background supply that dominates wherever photoionisation falls below it. The equilibrium solution of equation~(\ref{eq:ode}) recovers equation~(\ref{eq:nu_c_t}) exactly, including the noon normalisation. For a
non-synchronous rotator, each surface parcel carries its ionisation history
through the day–night cycle. Writing $x \equiv n_e/n_e^{\rm noon}$ and $\eta \equiv |\Omega| t$ for the local-time angle, with $\Omega$ the solar-day angular frequency of Section~\ref{sec:localcut}, equation~(\ref{eq:ode}) becomes the non-dimensional equation that we actually integrate,
\[
\frac{{\rm d}x}{{\rm d}\eta} \;=\; \frac{1}{2\pi\,\varepsilon}\Bigl[\max(\cos\chi,\,\xi) - x^{2}\Bigr],
\qquad
\varepsilon \equiv \frac{\tau_{\rm eff}}{P_{\rm sol}},
\qquad
\tau_{\rm eff} \equiv \frac{1}{\alpha_{\rm eff}\,n_e^{\rm noon}},
\]
with $\nu_c^{\rm loc}/\nu_c^{\rm noon} = x^{1/2}$. Two points about this equation should be made explicit. First, the relevant forcing period is the solar-day period $P_{\rm sol} = 2\pi/|\Omega_{\rm rot} - 2\pi/P_{\rm orb}|$ of Section~\ref{sec:localcut}, not in general the inertial rotation period; the two coincide for Earth but can differ greatly for close-in, resonant, or nearly synchronous planets. Secondly, the coefficient $\alpha_{\rm eff}$ that sets the relaxation time is \emph{not} the laboratory dissociative-recombination coefficient $\alpha$ of Section~\ref{sec:exo}: it is an empirical effective coefficient, chosen to represent the range of observed ionospheric relaxation times. With the laboratory value, equation~(\ref{eq:tau_rec}) gives a relaxation time of seconds, which is a lower bound (Section~\ref{sec:flares}). The observed relaxation of Earth's ionosphere is far slower, reflecting the rate-limiting ion--neutral chemistry and transport discussed there: effective response times range from tens of minutes in the lower F region to of order one to a few hours near the F2 peak \citep{Schunk2009}. We therefore present two illustrative cases that bracket this range for a 1-d solar day: a fast response, $\tau_{\rm eff} \approx 36$~min ($\varepsilon = 0.025$), and a slow response, $\tau_{\rm eff} \approx 2.4$~h ($\varepsilon = 0.1$), the latter being more representative of Earth's F2 peak. Larger values of $\varepsilon$ also describe planets with shorter solar days.

\begin{figure}
\centering
\includegraphics[width=\columnwidth]{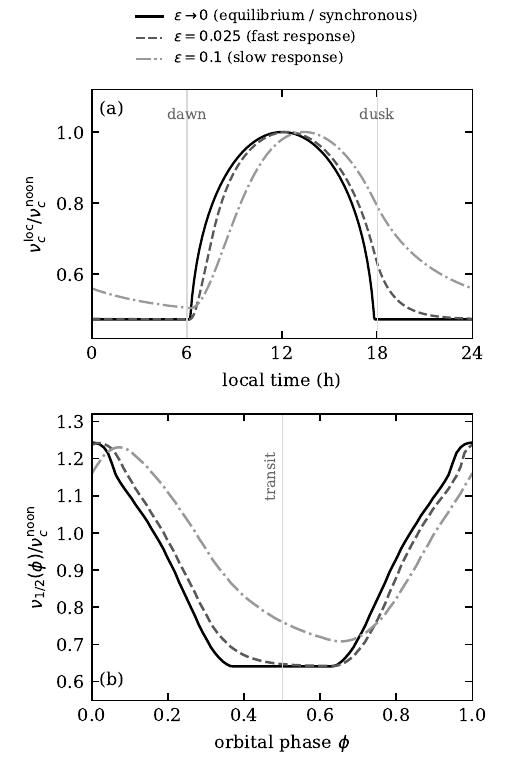}
\caption{Time-dependent generalisation of the cut-off model, from the non-dimensional form of equation~(\ref{eq:ode}) (Section~\ref{sec:ode}), for a solar day $P_{\rm sol} = 1$~d and prograde rotation. Three cases are shown: the equilibrium limit $\varepsilon \to 0$ (solid), which also describes synchronously rotating planets; a fast response, $\varepsilon = \tau_{\rm eff}/P_{\rm sol} = 0.025$ ($\tau_{\rm eff} \approx 36$~min; dashed); and a slow response, $\varepsilon = 0.1$ ($\tau_{\rm eff} \approx 2.4$~h; dot--dashed), closer to the behaviour of Earth's F2 peak.
\textit{Panel (a):} the steady-state local cut-off at the equator as a function of local time. A finite response time carries near-daytime densities past dusk, so that the nightside minimum is reached only just before dawn.
\textit{Panel (b):} the corresponding half-escape frequency $\nu_{1/2}(\phi)$ as a function of orbital phase, computed with the fiducial obliquity factor and Lambertian emission weighting of Fig.~\ref{fig:modulation}. The equilibrium curve is symmetric about transit. A finite response displaces the curve in the direction of rotation: the maximum lags superior conjunction by $\Delta\phi \approx 0.02$ (fast) and $0.07$ (slow), and the centre of the nightside plateau lags transit by $\approx 0.05$ and $0.15$ respectively. The fractional modulation $A = (\nu_{1/2}^{\max} - \nu_{1/2}^{\min})/\nu_{1/2}^{\max}$ is 48 per cent for the equilibrium and fast-response cases and 42 per cent for the slow response. In all cases the period of the modulation remains the orbital period.}
\label{fig:ode}
\end{figure}

Figure~\ref{fig:ode}(a) shows the steady-state local cut-off around a full
day, obtained by integrating equation~(\ref{eq:ode}) to a periodic state.
For finite $\varepsilon$, the cut-off no longer drops to the floor at the
dusk terminator: longitudes recently in sunlight retain near-daytime
densities and decay through the night, so the floor is reached only by the
longitudes that have been dark longest - the pre-dawn limb that is about to rotate back into starlight. The nightside minimum is thereby localised near dawn
rather than distributed uniformly. We note that this lagged pattern is still
steady in the frame co-rotating with the planet-star line, so the
conclusion of Section~\ref{sec:localcut} still stands: the disc-integrated modulation remains locked to the orbital phase.

The observable consequence, shown in Figure~\ref{fig:ode}(b), is a
distortion of the phase curve rather than a change in its period. Panel (b) shows the half-escape frequency $\nu_{1/2}(\phi)$ computed from equation~(\ref{eq:fesc}) with the lagged local pattern and the fiducial obliquity factor and emission weighting of Section~\ref{sec:discavg}. For the fast response ($\varepsilon = 0.025$) the fractional modulation is unchanged at $A = 48$ per cent, but the whole curve is displaced in the direction of rotation: the maximum lags superior conjunction by $\Delta\phi \approx 0.02$, and the centre of the nightside plateau lags transit by $\approx 0.05$. The slow response ($\varepsilon = 0.1$), closer to Earth's F2 peak, gives lags of $\approx 0.07$ and $\approx 0.15$ respectively, with $A$ reduced to 42 per cent. The lag thus constrains the effective atmospheric response time relative to the solar-day period. It is not a clean measurement of ion chemistry alone -- transport, winds, and plasma drifts all contribute to $\tau_{\rm eff}$, and additional phase shifts can be introduced by a non-uniform source distribution or by eccentric illumination -- but, with those caveats, its magnitude would bound the atmospheric response timescale independently of the flare response of Section~\ref{sec:flares}, and its sign carries information about the sense of rotation. For synchronously rotating
planets, no surface parcel crosses the terminator: in this idealised circular, zero-obliquity model the forcing is time-independent, the rotational lag vanishes, and the phase curve is symmetric about the transit phase.

\subsection{Flares as a competing source of time variability}
\label{sec:flarevar}

The phase-locked signal described above is not the only source of time variability in the cut-off frequency. As quantified in Section~\ref{sec:flares}, stellar flares from M-dwarf hosts can transiently enhance the XUV flux at the planet and drive the cut-off upwards on timescales of minutes to hours, although the $Q_\infty^{1/4}$ scaling limits even a $100\times$ XUV enhancement to a factor of $\sim 3.2$ in $\nu_c$, and bursts shorter than the atmospheric response time (Section~\ref{sec:flares}) produce only a partial response.

The two sources of variability can be separated, but not trivially. In the time-series analysis described in Section~\ref{sec:fold}, flare-driven enhancements appear as aperiodic positive excursions in $\nu_c$ that are uncorrelated with the planet's orbital period. Standard periodograms, including Lomb--Scargle, are nevertheless sensitive to large non-periodic excursions: flares raise the effective noise floor, can generate aliases through irregular sampling, and can bias the recovered amplitude and phase of the orbital signal. In practice, flare intervals should be identified through simultaneous X-ray or optical monitoring of the host star and then masked, explicitly modelled, or down-weighted, rather than assumed harmless.

\subsection{Confounding sources}
\label{conf}

Several natural processes produce low-frequency turnovers in radio
spectra, and several are also periodically variable. Any
claim of a technological origin for a phase-locked turnover must therefore
survive comparison with each of the following.

\textit{Stellar magnetic activity.} Active stars display rotationally
modulated radio emission tied to the stellar magnetic geometry \citep{Guedel2002}. This
variability, however, is locked to the \textit{stellar} rotation period, not
the planetary orbital period, and is accompanied by correlated
activity indicators (X-ray, optical flaring, chromospheric emission) that can
be monitored simultaneously. Such multiwavelength indicators are valuable but not definitive discriminants: the radio--X-ray relation carries substantial scatter and behaves differently for coherent and incoherent emission. The cleanest separation is by period, with near-commensurate stellar rotation and planetary orbital periods remaining the most difficult case.

\textit{Auroral cyclotron-maser instability (CMI) emission.} Planetary
magnetospheres are genuine periodic radio sources: Jupiter's decametric
emission is strongly modulated at the planetary rotation period
\citep{Zarka1998}, and analogous emission from exoplanets, modulated on rotational or orbital timescales, is an active area of research. CMI emission,
however, differs from an ionospheric turnover in almost every other respect:
it is generated at the local electron \textit{gyro}frequency rather than the
plasma frequency, is strongly circularly polarised (up to $\sim$100~per~cent), and is typically bursty and narrowly beamed. CMI can nevertheless produce drifting spectral structures, frequency-dependent visibility, and broad time-averaged emission, and propagation effects or unresolved averaging can reduce the observed circular polarisation, so no single CMI characteristic is decisive on its own. The fundamental distinction is that CMI \emph{generates} radiation, with spectral and temporal structure tied to the emission process itself, whereas the effect proposed here is a \emph{filter} imposed on pre-existing broadband radiation, with structure tied geometrically to the planet's day--night pattern.

\textit{Star--planet interaction (SPI).} Magnetic SPI can produce radio
emission modulated at the planetary orbital period or the star--planet
synodic (beat) period \citep{Zarka2007}. Where SPI emission is CMI-driven it shares the polarisation and burstiness discriminants above, but not every proposed SPI signal need be CMI-dominated: incoherent gyrosynchrotron or other magnetic emission may also contribute. SPI is in this sense the most demanding natural contaminant, because orbital phase coherence alone cannot distinguish it from the proposed ionospheric signature. The discrimination must instead rest on spectral character (a drifting turnover \emph{frequency} rather than a flux modulation), polarisation, and the location of the emission, which for SPI lies in the stellar corona rather than at the planetary position -- a difference that future high-resolution astrometry can, in principle, test.

\textit{Free--free absorption in ionised winds.} A dense stellar wind
imprints a low-frequency free--free turnover on any embedded source \citep{WrightBarlow1975}. The
turnover frequency, however, evolves on wind timescales and is not
phase-locked to the planetary ephemeris; the discriminant here is spectral shape and phase behaviour, not periodicity alone.

\textit{Gyrosynchrotron emission from stellar coronae.} Coronal gyrosynchrotron is an intrinsically generated, broad, magnetically controlled stellar component whose self-absorption turnover varies with the source opacity and the energetic-electron population \citep{Dulk1985}. It can be modulated as active coronal structures rotate through view, and in close interacting systems may even display orbital modulation, so it is not necessarily aperiodic on planetary timescales. The distinction is again generative versus filtering: a gyrosynchrotron turnover tracks the evolving stellar source itself, whereas the ionospheric turnover is an attenuation imposed on a pre-existing planetary-surface spectrum, phase-locked to the planet's day--night geometry and positionally coincident with the planet.

\textit{Circumplanetary plasma tori and intervening structures.} Absorption
by a plasma torus \citep[an Io-like configuration; e.g.][]{KaiserDeschBrown1996} could, in principle, be modulated
on the orbital period and is arguably the most awkward confounder. A torus can impose not only broadband absorption but also its own plasma-frequency cut-off, free--free absorption, refraction, and scattering, with mode-dependent propagation; and its electron density is spatially structured, as radio-occultation measurements of the Io torus demonstrate \citep{2018JGRA..123.6207P}. Torus
effects, however, track the torus geometry and density structure, and primarily modulate the \textit{amplitude} of the
received flux across a broad band, whereas the ionospheric diagnostic
predicts a shift of the turnover \textit{frequency} itself, following the
specific $(\cos\chi)^{1/4}$ law of equation~(\ref{eq:nu_c_t}) with its maximum locked to the substellar phase.

We therefore do not claim that periodic low-frequency variability is, by itself, unambiguous evidence of technology. It is not: the literature on planetary and stellar radio emission provides many natural examples of such variability. The value of the ionospheric diagnostic lies instead in the particular observational pattern expected for planetary-scale leakage: (i) a smooth spectral turnover rather than a beamed burst; (ii) weak polarisation, in contrast to CMI processes; (iii) a characteristic \textit{frequency}, not merely an amplitude, that varies with phase; (iv) modulation following the $(\cos\chi)^{1/4}$ Chapman scaling, with $\nu_c^{\rm noon}$ as the principal frequency scale and a small set of geometric and atmospheric nuisance parameters; and (v) phase-locking to an independently determined orbital ephemeris, with the maximum cut-off occurring at the substellar phase or lagging it by the predictable amount set by the atmospheric response (Section~\ref{sec:ode}). Each of these properties has natural analogues when considered in isolation; their combination, to the best of our knowledge, does not. A low-frequency turnover at the predicted frequency is thus the basic signature of a surface-origin BRaT (Section~\ref{sec:static}), but it is not by itself a sufficient one. It is the orbital-phase behaviour of that turnover which, combined with the associated broadband continuum detection, would substantially strengthen a technological interpretation.

\section{Detection prospects}
\label{sec:obs}

BRaTs are expected to be faint $\mu$Jy and sub-$\mu$Jy radio sources \citep{2026MNRAS.549ag991G}. Detecting an ionospheric cut-off is therefore considerably more challenging than simply detecting the continuum emission itself. In this section, we outline a possible long-term detection strategy and discuss the main observational challenges.

\subsection{Frequency range}

The most efficient frequencies for the initial detection of BRaTs are likely to lie above $\sim 1$~GHz, where the sky noise is relatively low, the planetary ionospheres are essentially transparent, and the ionospheric cut-off effects are negligible. At these frequencies, facilities such as SKA-Mid and the ngVLA are expected to provide $\mu$Jy and sub-$\mu$Jy continuum sensitivities over broad instantaneous bandwidths \citep{Braun2019,Selina2018}, making them natural instruments for the initial discovery of candidate BRaT systems.

A candidate identified at GHz frequencies would naturally be followed up at lower frequencies, approaching the ionospheric cut-off values predicted for exoplanets orbiting Sun-like stars and M-dwarfs. From Table~\ref{tab:exo}, the expected cut-offs for most systems lie in the $5$–$30$~MHz range, rising to $\sim 40$--$120$~MHz only for planets around active M-dwarfs (Section~\ref{sec:exo}). The first objective of such follow-up is simply to establish whether the continuum turns over near the predicted frequency (Section~\ref{sec:static}); this needs only a time-averaged, multi-band spectrum. The phase-resolved measurements of Section~\ref{sec:fold} are a second, more demanding step. The principal range of interest, $\sim 5$–$50$~MHz, is therefore only partly accessible to planned large facilities such as SKA-Low, whose nominal frequency coverage is $50$–$350$~MHz. Ironically, the lowest frequencies are inaccessible from the ground due to the same phenomenon considered here: Earth’s ionosphere imposes a cut-off near $\sim 10$~MHz and severely compromises calibration at frequencies below a few tens of MHz, where phase coherence times become very short. Facilities such as LOFAR and the LWA can observe below $\sim 50$~MHz (the MWA band, by contrast, begins near $\sim70$~MHz), although with less sensitivity than SKA-Low will provide within its own band. Nevertheless, detections of bright sources, including pulsars \citep{2016A&A...586A..92P} at frequencies as low as 15~MHz are encouraging.

\subsection{Sensitivity and cadence}

At the lower edge of the SKA-Low band ($\sim$50--60~MHz), conditions are far harsher than at the band centre. The originally projected full-station thermal sensitivity is $\sim 160~\mu{\rm Jy~beam^{-1}}$ in 1~hr \citep{Braun2019}, driven primarily by the steeply rising Galactic sky temperature, and the current SKAO performance estimates additionally quote a classical confusion limit of $\sim 680~\mu{\rm Jy~beam^{-1}}$ at these frequencies \citep{SKAO2021}. Extracting weak signals at the lowest frequencies will therefore require the longest-baseline configurations, positional prior knowledge of the target, and careful subtraction of confusing sources. The orbital-phase modulation of Section~\ref{sec:diurnal} is therefore unlikely to be detectable in individual observations, except perhaps for the nearest and brightest BRaT candidates. The observational objective would be to recover a weak, phase-coherent modulation by combining data over many orbital cycles. For an especially compelling BRaT candidate, this could justify a dedicated low-frequency campaign with a total integration time of many weeks, perhaps months. The star-anchored nature of the pattern (Section~\ref{sec:localcut}) simplifies the observing cadence considerably. Because the apparent spectrum varies only slowly with orbital phase, individual integrations lasting a substantial fraction of an orbital phase bin can be accumulated with little smearing: a one-day integration on a ten-day orbit averages over $\Delta\phi = 0.1$, which for a near-sinusoidal modulation costs only $\approx 2$~per~cent of the amplitude. The tolerable integration length should be determined from the adopted phase template;  there is, in particular, no need to resolve the planet's rotation. Conversely, integrations spanning one or more rotation periods deliberately average out any secondary modulation arising from a very uneven geographic distribution of surface transmitters (see Section~\ref{sec:caveats}). We stress that this isolates the geometrically determined component of the modulation only under the explicit assumption that the leakage is sufficiently widespread in longitude and statistically stationary over the campaign; the recovered spectrum always retains its dependence on the transmitters' luminosity, spectrum, and duty cycle.

\subsection{Phase-folded detection}
\label{sec:fold}

The appropriate search strategy is a matched spectral--phase template search, analogous in a general sense to periodic-signal searches in transit photometry \citep{Kovacs2002} and pulsar astronomy. Rather than attempting to measure a cut-off frequency in each observation, the data are folded on a trial orbital period and phase.

For a trial period $P$ and reference epoch $t_0$, each observation is assigned an orbital phase,
\begin{equation}
\phi(t)=\left(\frac{t-t_0}{P}\right)\ {\rm mod}\ 1.
\end{equation}

Spectra obtained at similar orbital phases are then combined. If the assumed period is incorrect, the modulation is smeared out, and the folded spectrum approaches the time-averaged continuum. If the correct period is used, the ionospheric modulation adds coherently while random noise averages down. 
A genuine signal will manifest as a single peak centred on the phase
at which the substellar hemisphere faces the observer -- or displaced from it by the lag expected for a finite atmospheric response (Section~\ref{sec:ode}), which forms part of the template rather than a discrepancy -- with a fractional modulation $A = (\nu_{1/2}^{\max} - \nu_{1/2}^{\min})/\nu_{1/2}^{\max} \approx 48$~per~cent in the fiducial model ($39$--$48$ per cent across the emission models of Section~\ref{sec:discavg}). The light curve remains within the upper 20 per cent of its range for only $\approx 0.16\,P$, and sits on its nightside plateau for $\approx 0.25\,P$ around transit, reflecting the rapid hand-over from dayside to nightside dominance
near quadrature (Section~\ref{sec:discavg}). Quantitative agreement
between the observed light curve and a Chapman-model template with
$\nu_c^{\rm noon}$ as the principal frequency scale -- together with a small
set of geometric and atmospheric nuisance parameters (the inclination, the
night-side floor $\xi$, the response parameter $\varepsilon$, the emission weighting and obliquity factor of Section~\ref{sec:discavg}, and the
transmission-profile width) -- provides a
constrained, predictive test that few astrophysical alternatives can pass.

Panel (a) of Figure~\ref{fig:modulation} is, in effect, the template light curve against which folded data would be compared.
The search is therefore fundamentally a model-comparison problem: one tests
whether the phase-dependent escape-fraction model $f_{\rm esc}(\nu,\phi)$ of
equation~(\ref{eq:fesc}), with the Chapman template of
Section~\ref{sec:diurnal} and $\nu_c^{\rm noon}$ as its principal frequency scale, provides a significantly better description of the accumulated data than a static spectrum.
Where the orbital period is already known from transit photometry, radial-velocity, or astrometric observations, as it might well be for any credible BRaT candidate, both the expected period and phase are fixed in advance, which greatly reduces the trials factor. The analysis nevertheless remains a constrained template-fitting or model-comparison problem rather than a literal single-hypothesis test, since the atmospheric, geometric, continuum, and calibration parameters are still fitted. The signature is thus not a single spectral measurement but a coherent time-domain signal recovered through long-term integration. At frequencies where the array is not confusion-limited (Section~\ref{sec:obs}), coherent integration times of several months could, in principle, approach $\mu$Jy sensitivity levels, making outstanding BRaT candidates possible targets.

\subsection{Verification}
\label{sec:verif}

A convincing detection must satisfy several independent criteria, which together define how a false positive would be recognised and rejected:

\begin{enumerate}[labelwidth=!,align=left]
    \item \textit{Phase coherence.} The folded signal must strengthen as additional data are added, remain stable across independent observing campaigns, and lose coherence, or decrease substantially in significance, when the data are folded at incorrect trial periods (aliases and the observing-window function can leave residual peaks, so the signal need not vanish entirely).
    \item \textit{Ephemeris agreement.} The recovered period and phase must
    match the independently determined orbital ephemeris, with the maximum apparent cut-off occurring at superior conjunction in the instantaneous-equilibrium model, or displaced from it by the phase lag expected for a finite atmospheric response (Section~\ref{sec:ode}); the comparison must be made against the appropriate equilibrium or lagged template. For a transiting system the planet is occulted at exact superior conjunction, so the full-dayside spectrum is sampled immediately before and after secondary eclipse.
    \item \textit{Spectral localisation.} The variability must be confined
    to frequencies near the spectral turnover, with the continuum well above the cut-off remaining steady in phase.
    \item \textit{Weak polarisation.} The modulated component should show
    the weak polarisation expected of incoherent broadband leakage, in
    contrast to the strong circular polarisation of auroral CMI emission
    (Section~\ref{conf}).
    \item \textit{Astrometric association.} The emission must be positionally
    coincident with the planetary system, sharing its proper motion and
    parallax.
    \item \textit{Independence from stellar activity.} Simultaneous X-ray
    and optical monitoring of the host star must show no correlation between
    the phase-locked modulation and stellar flaring. A recovered period distinct from the stellar rotation period is a strong discriminant where available, but not a necessary condition: if the two periods are equal, nearly equal, or commensurate, discrimination must rely instead on spectral morphology, multi-wavelength activity indicators, polarisation, and comparison with stellar-activity and star--planet-interaction models (Section~\ref{conf}).
    \item \textit{Full template agreement.} Beyond phase coherence at
    the correct period, the signal must match the complete predicted
    spectral--phase template: the migration of $\nu_{1/2}$ with phase, the
    phase-dependent turnover width (Section~\ref{sec:discavg}), and
    repeatability of the pattern across independent frequency sub-bands.
    \item \textit{Terrestrial-interference rejection.} Radio-frequency
    interference must be excluded explicitly: through simultaneous
    off-source reference beams, interferometric localisation of the signal
    to the target position, repetition across epochs and (where possible)
    across geographically separated stations, and standard instrumental and
    environmental checks.
\end{enumerate}

No single criterion would be decisive on its own. However, a candidate satisfying all eight criteria would be difficult to reconcile with the main false-positive classes discussed above: terrestrial interference, stellar activity, auroral CMI emission, or unrelated background sources. In particular, a weakly polarised spectral turnover that is astrometrically associated with the planetary system and phase-locked to the independently known orbital ephemeris would elevate the source to a high-priority technosignature candidate warranting sustained follow-up.

The main practical limitations are access to the appropriate frequency range and good sensitivity. A proper exploration of the $\lesssim 30$~MHz regime may therefore require a major space-based or lunar-farside low-frequency radio observatory. Such facilities on the required scale are unlikely to be available in the near term, making the ionospheric cut-off a challenging diagnostic for present instruments but a potentially powerful target for future low-frequency radio astronomy.

\section{Discussion}
\label{sec:discussion}

\subsection{The cut-off as a planetary atmospheric diagnostic}

Although motivated here as a technosignature discriminator, the phase-locked cut-off is, in the first instance, a measurement of planetary atmospheric physics. A detection would directly constrain the peak electron density of the ionosphere through equations~(\ref{eq:nu_p}) and (\ref{eq:nu_c_t}). It would also constrain the day-night electron density contrast through the modulation amplitude. However, the amplitude is degenerate with the orbital inclination, the source distribution, the angular emission pattern of the leakage, and the adopted transmission profile, so the empirical floor $\xi$ must be inferred jointly with these parameters rather than measured directly. The observations would further constrain the nightside plasma transport and residual ionisation that physically set this floor. Finally, the effective atmospheric response time, $\tau_{\rm eff}$, could be inferred from the dusk--dawn lag of the phase curve (Section~\ref{sec:ode}). The sign of the lag additionally carries information about the sense of rotation, subject to the caveats of Section~\ref{sec:ode}. Ultraviolet transit spectroscopy, atmospheric-escape signatures, auroral radio emission, and star--planet plasma interactions have all been discussed as indirect probes of ionised upper atmospheres; the approach proposed here is unusual in offering a direct constraint on the peak ionospheric plasma frequency of a terrestrial exoplanet and on its variation in local time. The observational programme outlined in Section~\ref{sec:obs} is admittedly long-term but does have scientific value beyond the field of technosignatures. 


\subsection{Value for SETI}

The diagnostic has a property that distinguishes it from most proposed
technosignatures: it depends on planetary physics, not on the behaviour of
the transmitting civilisation. Most radio SETI strategies must assume
something about intent—deliberate beacons, particular frequency choices,
persistence, or duty cycle. The ionospheric imprint requires none of this. Any surface-generated broadband leakage, whatever its purpose, spectrum or
schedule, must cross the ionosphere on its way out, and acquires the
phase-locked turnover in doing so. The ionospheric transfer function itself is set by planetary physics, and its phase is computable in advance from the system's known geometry. Recovering it observationally does, however, still make demands of the transmitters: the leakage must contain appreciable power across the cut-off, escape upward, persist long enough for phase folding, and be reasonably stationary (Sections~\ref{sec:obs} and \ref{sec:caveats}). What the diagnostic removes is any assumption about \emph{intent}; it does not remove the need for a detectable, persistent transmitter. We note also that a photoionised ionosphere is a common consequence of any irradiated atmosphere -- it is neither unique to nor guaranteed by habitability -- so the cut-off converts a near-universal feature of irradiated planetary atmospheres, rather than of habitability as such, into a verification tool.

\subsection{Caveats}
\label{sec:caveats}

Several simplifications in our treatment should be kept in view. The
Chapman description assumes photochemical equilibrium with a single dominant ion species; real ionospheres have multiple layers, and transport becomes important at high altitudes and on the nightside, where our phenomenological floor $\xi$ absorbs a good deal of unmodelled physics (the time-dependent treatment of Section~\ref{sec:ode} relaxes the equilibrium assumption but retains the single-species chemistry). 

A second point concerns the geographic distribution of the leakage across the planet's surface. Our disc-averaged cut-off behaves like a smooth,
area-weighted turnover only when many longitudes contribute to the integrated spectrum. The turnover is in every case a property of the ionosphere rather than of the source. Any radiation that escapes must first pass through the ionosphere above its own emission region, whose cut-off is fixed by the local stellar illumination and hence by local time, but the source geography determines \emph{which} ionosphere dominates the average. If the leakage came instead from a handful of localised transmitters, the integrated spectrum would be weighted
towards the ionosphere above those sites, so that the phase and amplitude of the turnover modulation would track their particular longitudes rather than the disc as a whole. Broad longitudinal coverage must therefore remain an explicit model assumption. It is tempting to argue that a Kardashev Type~I civilisation, harnessing energy on a planetary scale, would spread its leakage across all longitudes, but planetary-scale energy use implies nothing definite about transmitter geography, spectrum, beaming, or the adoption of radio-quiet technologies. Earth at least provides an existence proof: its unintended radio emission, though
clustered on land \citep{2023MNRAS.522.2393S}, spans the full range of
longitudes, so that the disc-integrated terrestrial signal is nothing like being dominated by a single longitude. In this broad-coverage limit, the turnover frequency and any accompanying flux modulation are complementary observables with different sensitivities: both depend on the ionospheric transmission and on the source geography, but the turnover frequency responds primarily to the planetary ionosphere, while the broadband flux responds primarily to the geographic distribution and duty cycle of the emitters.

Finally, our scaling relations assume that the leakage spectrum is intrinsically broadband down to the cut-off; a civilisation whose emission is concentrated well above $\nu_c$ would carry no measurable ionospheric imprint. None of these caveats alters the qualitative conclusion, but each would need to be revisited
in the interpretation of any real candidate.

\section{Conclusions}
\label{sec:conc}

Any radio leakage generated on the surface of a planet possessing a sufficiently ionised atmosphere -- a common consequence of stellar irradiation, neither unique to nor guaranteed by habitability -- must propagate through the planet’s own ionosphere before escaping into space. The ionosphere, therefore, imposes a fundamental low-frequency boundary on broadband radio technosignatures (BRaTs), analogous to the role played by the Earth’s ionosphere in terrestrial radio astronomy. In this paper, we have developed the ionospheric cut-off both as a physical constraint on the detectability of planetary-surface radio emission and as a potential technosignature in its own right.

Using a simple Chapman-layer description of photoionised planetary atmospheres, we have estimated the ionospheric cut-off frequencies expected for terrestrial planets in a range of habitable-zone environments. Because the peak electron density is set by a balance between photoionisation and recombination, the cut-off frequency depends only weakly on the incident stellar XUV flux, scaling as $Q_{\infty}^{1/4}$. Consequently, even large differences in stellar activity translate into relatively modest changes in the cut-off frequency. For planets orbiting Sun-like stars and many K-dwarfs, the cut-off is expected to remain broadly comparable to Earth’s value, while planets in the habitable zones of active M-dwarfs may exhibit cut-offs extending into the tens of MHz, potentially rising towards $\sim100$~MHz during strong, sustained flares or in the most extreme irradiation environments.

We have also shown that the ionospheric cut-off should not be regarded as a perfectly sharp spectral boundary. Oblique propagation through the ionosphere, magnetised plasma effects, and collisional absorption all act to broaden the transition between trapped and escaping radiation. The observable signature is therefore expected to be a smooth low-frequency turnover rather than an abrupt spectral edge, with the detailed shape depending on the structure and composition of the planetary ionosphere. This turnover is the primary signature. It is present at every orbital phase and for every viewing geometry, at a characteristic frequency that can be estimated from the XUV output of the host star, and it can be sought in a time-averaged spectrum without an orbital ephemeris. Because natural low-frequency turnovers exist, it is a necessary rather than a sufficient condition for a surface origin; but continuum emission persisting well below the predicted plasma frequency would be very difficult to reconcile with one.

The second main result of this paper is that the cut-off need not be static. Through the Chapman scaling, the local cut-off varies approximately as $(\cos\chi)^{1/4}$, reaching its maximum at the sub-stellar point and declining towards a nightside floor maintained by residual ionisation and plasma transport. This day--night pattern is anchored to the substellar point rather than to the rotating surface, so to a distant observer, the disc-averaged cut-off is modulated on the planet's \emph{orbital} period, sweeping between a dayside maximum at superior conjunction and the nightside floor near the transit phase. Including the dependence of the escape threshold on the emission angle towards the observer raises the absolute level of the turnover by $\approx 40$ per cent and broadens it, but leaves its fractional modulation essentially unchanged: the half-escape frequency varies by $A \approx 48$ per cent over the orbit in the edge-on, Earth-calibrated fiducial model ($39$--$48$ per cent across plausible angular emission patterns of the leakage), with a phase set by planetary geometry; the amplitude additionally depends on the orbital inclination and on the spatial distribution of the sources. Departures from photochemical equilibrium displace the phase curve by an amount that constrains an effective atmospheric response timescale relative to the solar-day period, without altering the period of the modulation itself.

This behaviour suggests a new potential diagnostic for planetary-surface radio emission. A broadband radio source associated with an exoplanet that exhibits a low-frequency spectral turnover, whose characteristic frequency varies periodically with the planet’s known orbital period, would be difficult to explain through conventional astrophysical processes. While individual natural processes can reproduce a low-frequency
turnover, and some can reproduce periodic variability, we are not aware of any known natural process that reproduces the specific combination of a smooth, weakly polarised turnover, a stable modulation of the turnover frequency itself following the
$(\cos\chi)^{1/4}$ Chapman scaling, and the geometric locking of the
maximum cut-off to the planet's substellar phase (up to a predictable atmospheric lag)
(Section~\ref{conf}).

Detecting such a signal will be challenging, and we emphasise that the cut-off is best interpreted alongside other evidence: a detection satisfying the verification criteria of Section~\ref{sec:verif} would provide
strong supporting evidence, though not by itself conclusive proof of an artificial planetary-surface origin. BRaTs are expected to be intrinsically faint, and the recovery of the modulation will likely require long integrations accumulated over many orbital cycles. The cut-off can be recovered by phase-folding many thousands of hours of data at the planet’s known orbital period, in a matched spectral--phase template search analogous in spirit to box-least-squares transit searches. A two-pronged programme may therefore be envisaged: detection-optimised BRaTs continuum surveys at $\gtrsim1$~GHz with facilities such as SKA-Mid and the ngVLA, followed by low-frequency multi-epoch monitoring with LOFAR (10--90~MHz and 110--240~MHz; \citealt{vanHaarlem2013}) and SKA-Low (50--350~MHz; \citealt{Mellema2019}), whose bands together span the relevant range, to search for the phase-locked diagnostic. Access to the lowest frequencies, below $\sim30$~MHz, is fundamentally limited from the ground by Earth’s own ionosphere — the same physical effect that the diagnostic itself exploits. Future space-based or lunar-farside observatories \citep{Burns2021} would therefore provide a natural extension of the programme advocated here.

The diagnostic developed here will not be tested quickly. It requires low-frequency access and sensitivities that only future facilities can provide, together with sustained multi-epoch monitoring, ancillary host-star data, and -- as the caveats above emphasise -- favourable properties of the transmitting source itself. The history of SETI counsels patience and a willingness to revisit assumptions as new data emerge. If a planetary-surface technological signal is ever identified at radio wavelengths, the imprint of the planet's own ionosphere, a filter imposed by ordinary atmospheric physics on any escaping transmission, may prove to be among the most distinctive clues to its origin.

\section*{Acknowledgements}

The author is grateful to the anonymous referee for a very careful and expert review of this paper. Their detailed comments on ionospheric propagation, plasma chemistry, and the limitations of the disc-averaged model corrected several errors, sharpened the physical arguments, and substantially improved the manuscript overall. 
Generative artificial intelligence tools were used in the preparation of this work. \citep{Claude} was used in the development of the Python code for the disc-integrated escape model and the production of the figures, and with checking calculations. The concept, the scientific content, the analysis choices and the conclusions are the author's own, and the author has reviewed and verified all AI-assisted text, code and numerical results, and takes full responsibility for the content of the paper.


\section*{Data Availability}

No new observational data were generated or analysed in support of this research. The Python code used to generate Figures~1--4, including the disc-integrated escape-fraction model of Section~\ref{sec:discavg}, is available from the author on request.


\bibliographystyle{mnras}
\bibliography{references_new_fixed} 


\bsp	
\label{lastpage}
\end{document}